\documentclass[twocolumn, journal]{IEEEtran}
\IEEEoverridecommandlockouts

\usepackage{color}
\usepackage{xcolor}
\usepackage{bm}
\usepackage{graphicx}
\usepackage{amsmath}
\usepackage{amssymb}
\usepackage{multirow}
\usepackage{algorithm}
\usepackage{algorithmic}
\usepackage{booktabs}
\usepackage{array}
\usepackage{amsthm}
\usepackage{lipsum}
\usepackage{enumerate}
\usepackage{stfloats}
\usepackage{cases}
\usepackage{cite}
\usepackage{mathtools}
\usepackage{cuted}
\usepackage[caption=false,font=footnotesize]{subfig}
\usepackage{geometry}

\def\C{\mathbb{C}}

\def\Y{\mathbf{Y}}

\allowdisplaybreaks

\newtheorem{remark}{Remark}

\newtheoremstyle{noparens}%
  {}{}%
  {\itshape}{}%
  {\bfseries}{.}%
  { }%
  {\thmname{#1}\thmnumber{ #2}\mdseries\thmnote{ #3}}

\theoremstyle{noparens}

\title{Frequency-Dependent Hybrid Digital and Microwave Linear Analog Computer (MiLAC) Beamforming for Multiuser MIMO-OFDM Systems}

\author{Yiyang Peng, Zheyu Wu,~\IEEEmembership{Member, IEEE}, and Bruno Clerckx,~\IEEEmembership{Fellow, IEEE}
\thanks{Yiyang Peng, Zheyu Wu and Bruno Clerckx are with the Department of Electrical and Electronic Engineering, Imperial College London, London SW7 2AZ, U.K. (email: \{yiyang.peng22, zheyu.wu, b.clerckx\}@imperial.ac.uk).
}}

\begin{document}
\maketitle
\thispagestyle{empty}

\begin{abstract}
Microwave linear analog computing (MiLAC) has recently emerged as a promising approach for reducing the hardware and digital-processing requirements of large-scale multiple-input multiple-output (MIMO) systems. In narrowband multiuser systems, hybrid digital-MiLAC beamforming can achieve digital beamforming flexibility with only as many radio-frequency (RF) chains as data streams. However, extending this advantage to wideband systems is nontrivial because practical MiLAC circuits exhibit frequency-dependent responses. In this paper, we investigate hybrid digital-MiLAC beamforming for multiuser multiple-input single-output (MU-MISO) orthogonal frequency division multiplexing (OFDM) systems with a frequency-dependent MiLAC circuit model. We characterize the frequency dependence of MiLAC by relating the susceptance of each tunable component across subcarriers to its center-frequency value. Under this frequency coupling, we show that a single MiLAC circuit generally cannot simultaneously match the desired beamforming directions across all subcarriers, highlighting the essential role of subcarrier-dependent digital precoding. We then develop a block coordinate descent (BCD) algorithm to jointly optimize the digital and MiLAC beamformers for average sum-rate maximization. In particular, an efficient alternating direction method of multipliers (ADMM) algorithm is custom-designed to handle the frequency-coupled circuit constraints in the MiLAC beamformer subproblem. Simulation results show that the proposed hybrid digital-MiLAC design achieves a favorable tradeoff between sum-rate performance and RF-chain number. It outperforms conventional hybrid digital-analog beamforming with the same number of RF chains, particularly when fewer RF chains are available and the channel is more frequency-selective. With $64$ transmit antennas and $4$ data streams, it attains $96.3\%$ of the digital sum-rate using only $8$ RF chains, corresponding to $12.5\%$ of the digital RF-chain requirement. Furthermore, under highly frequency-selective channels and using $4$ RF chains, hybrid digital-MiLAC beamforming provides a $48\%$ sum-rate gain over the conventional hybrid design.
\end{abstract}

\begin{IEEEkeywords}
Microwave linear analog computer (MiLAC), orthogonal frequency division multiplexing (OFDM), beamforming, hybrid digital-MiLAC architecture.
\end{IEEEkeywords}

\section{Introduction}
Future sixth-generation (6G) wireless communication systems are expected to support increasingly high data rates, massive connectivity, and ubiquitous coverage, motivating the use of wider bandwidths and large-scale antenna arrays in massive or gigantic multiple-input multiple-output (MIMO) systems \cite{6G}. However, conventional fully-digital beamforming requires a dedicated radio-frequency (RF) chain for each transmit antenna, leading to prohibitive hardware cost, power consumption, and signal processing complexity as the number of antennas grows. A widely studied alternative is hybrid digital-analog beamforming, where a low-dimensional digital precoder is followed by an analog beamformer typically implemented using a phase-shifter network \cite{hybrid_narrowband,hybrid_ofdm}. By employing substantially fewer RF chains, hybrid digital-analog beamforming alleviates the hardware burden of fully-digital architectures. Nevertheless, the phase-shifter-based analog beamformer is subject to structural constraints that limit its design flexibility. Moreover, the same analog beamformer is commonly shared across all frequencies \cite{hybrid_ofdm}, further limiting its adaptability to frequency-selective channels.

Recently, microwave linear analog computing (MiLAC) has emerged as a promising alternative for large-scale MIMO beamforming \cite{nerini_milac_partI,nerini_milac_partII}. A MiLAC is a reconfigurable multiport microwave network composed of tunable impedance/admittance components, which can implement linear transformations directly in the analog RF domain as signals propagate through the network. When used for transmit beamforming, the input and output ports of the MiLAC are connected to RF chains and transmit antennas, respectively \cite{nerini_milac_partII,nerini_milac_magazine}. By tuning the internal admittance components, the MiLAC directly shapes the transmit signals through microwave interactions. Hence, beamforming is performed by the microwave network itself rather than through symbol-level digital matrix-vector processing. This makes MiLAC particularly attractive for large-scale antenna arrays, as the required number of RF chains scales with the number of data streams, while shifting a substantial part of the beamforming operation from the digital domain to the microwave network.

Early studies of MiLAC considered an idealized model with arbitrarily tunable complex admittances, under which a MiLAC can realize arbitrary beamforming matrices \cite{nerini_milac_partI,nerini_milac_partII}. To account for more practical microwave-network implementations, subsequent studies incorporated lossless and reciprocal constraints into MiLAC-aided beamforming. In point-to-point MIMO systems, MiLAC-aided beamforming has been shown to achieve the same capacity as fully-digital beamforming \cite{nerini_milac_capacity}. The multiuser case is more challenging: a lossless and reciprocal MiLAC generally cannot reproduce arbitrary digital beamformers, although the performance gap decreases as the user channels become more orthogonal \cite{tianyu_milac_beamforming,zheyu_milac_beamforming}. To overcome this fundamental limitation, hybrid digital-MiLAC \cite{zheyu_milac_beamforming} and two-layer MiLAC \cite{xiaohua_two_layer} beamforming architectures were proposed, both of which can achieve fully-digital beamforming performance in multiuser systems using only as many RF chains as data streams. To further reduce the circuit complexity, stem-connected MiLAC architectures were developed \cite{nerini_milac_stem}, which can preserve the performance of fully-connected MiLAC while substantially reducing the number of tunable components \cite{nerini_milac_stem,yuchen_stem}. Recent studies have further extended MiLAC to channel estimation \cite{qiaosen_milac_ce,qiaosen_milac_ce_beamforming}, sensing \cite{ziang_sensing}, and energy-efficient design \cite{binggui_lossy_milac,yuchen_ee,marco_milac_ee}. Other works have investigated practical hardware effects, including finite-resolution components \cite{zheyu_milac_discrete}, lossy components \cite{binggui_lossy_milac}, mutual coupling \cite{nerini_milac_physics_compliant}, and impedance mismatch \cite{zhang_rui_impedance_mismatch}.

Although the advantages of MiLAC-aided beamforming have been widely demonstrated, most existing studies focus on narrowband systems. An initial extension to wideband systems was presented in \cite{wideband_fi_milac_yiyang}, where a hybrid digital-MiLAC architecture was studied under the assumption that the MiLAC beamformer remains the same over the entire bandwidth. This work showed that, unlike the narrowband case, using only as many RF chains as data streams is insufficient to realize arbitrary digital beamformers in wideband systems. However, the frequency-independent MiLAC model adopted in \cite{wideband_fi_milac_yiyang} may not accurately characterize practical wideband MiLAC systems, since the response of a microwave network generally varies with frequency \cite{pozar2009microwave,nerini_milac_magazine}. In particular, the frequency dependence of MiLAC mainly originates from the reactive components in its microwave network, such as tunable capacitors and inductors, whose admittances vary with frequency \cite{wideband_circuit_hongyu}. Consequently, a fixed MiLAC circuit generally produces different beamforming responses at different frequencies.

This frequency-dependent behavior fundamentally changes the beamforming characteristics of MiLAC in wideband systems. On the one hand, a fixed MiLAC circuit produces different beamforming responses across frequencies, which may help accommodate frequency-varying wireless channels. On the other hand, all frequency responses are jointly determined by the same physical MiLAC circuit parameters and cannot be independently adjusted at each frequency. Therefore, it remains unclear whether the inherent frequency variation of MiLAC can provide sufficient flexibility to realize the desired beamforming responses across the bandwidth in frequency-selective channels. This motivates investigating the wideband beamforming flexibility and achievable performance of a frequency-dependent MiLAC. Furthermore, hybrid digital-MiLAC beamforming architecture offers a promising approach to enhance this flexibility by introducing subcarrier-specific digital precoders to complement the frequency-dependent MiLAC response. However, the achievable performance of such a hybrid architecture compared with fully-digital and conventional hybrid digital-analog beamforming in wideband multiuser systems remains to be investigated.

Motivated by these questions, in this paper, we investigate MiLAC-aided beamforming architectures for multiuser multiple-input single-output (MU-MISO) orthogonal frequency division multiplexing (OFDM) systems by explicitly modeling the frequency-dependent response of the MiLAC circuit. The main contributions are summarized as follows.

1) \textit{Frequency-dependent wideband hybrid digital-MiLAC beamforming modeling:} We develop a hybrid digital-MiLAC beamforming model for MU-MISO-OFDM systems that explicitly accounts for the frequency-dependent MiLAC circuit response. The MiLAC is represented by an equivalent finite-memory time-domain approximation and cascaded with the multipath wireless channel, leading to an equivalent set of parallel frequency-domain subcarrier models. Since a MiLAC relies on a reconfigurable microwave network, implemented using components similar to those used in the reconfigurable impedance network of a beyond-diagonal reconfigurable intelligent surface (BD-RIS), we adopt the circuit-based frequency-dependent model of BD-RIS in \cite{wideband_circuit_hongyu} for the tunable admittance components, where the responses across subcarriers are jointly determined by center-frequency susceptances. 

2) \textit{Wideband MiLAC-aided beamforming flexibility and limitations:} We investigate the beamforming flexibility of frequency-dependent MiLAC and the impact of using one MiLAC circuit over the entire bandwidth. Using a single-user system with two transmit antennas as an illustrative example, we analyze the beamforming gain, magnitude efficiency, and directional alignment, and show the difficulty of simultaneously matching the desired beamforming directions across all subcarriers using one MiLAC circuit. These results reveal the limitations of wideband MiLAC-aided beamforming and motivate the use of a hybrid digital-MiLAC architecture.

3) \textit{Joint wideband hybrid digital-MiLAC beamforming design:} We formulate an average sum-rate maximization problem that jointly optimizes the digital and MiLAC beamformers under lossless, reciprocal, and frequency-dependent circuit constraints with a radiated-power budget. To solve this nonconvex problem, we reformulate it into an equivalent weighted minimum mean-square error (WMMSE) form and design a block coordinate descent (BCD) algorithm. In particular, we handle the frequency-coupled MiLAC beamformer subproblem using an alternating direction method of multipliers (ADMM) framework to tackle the nonlinear relationship between the shared circuit susceptances and the resulting beamforming matrices across subcarriers. For efficient implementation, we further exploit the problem structure to simplify the Lagrange multiplier search in the beamformer updates into scalar evaluations and reformulate the susceptance update as a box-constrained least-squares problem that can be efficiently solved by projected gradient descent (PGD).

4) \textit{Performance evaluation and practical considerations:} Simulation results demonstrate that the proposed hybrid digital-MiLAC design approaches fully-digital beamforming performance with substantially fewer RF chains, and its gain over conventional phase-shifter-based hybrid digital-analog beamforming becomes more pronounced under high channel frequency selectivity and small number of RF chains. With $64$ transmit antennas and $4$ transmit streams, the proposed design achieves $96.3\%$ of the fully-digital sum-rate using $12.5\%$ of the RF chains. Under highly frequency-selective channels, it further provides a $48\%$ sum-rate gain over conventional hybrid digital-analog beamforming with only $4$ RF chains. We further characterize the effective temporal memory introduced by the frequency-dependent MiLAC and show that the resulting OFDM cyclic prefix (CP) overhead can be as low as $9.22\%$.

\textit{Organization:} Section~\ref{sec:system_model_hybrid} presents the wideband hybrid digital-MiLAC system model and the frequency-dependent circuit model of MiLAC. Section~\ref{sec:wideband_milac_flexibility} investigates the wideband MiLAC-aided beamforming flexibility and limitations. Section~\ref{sec:optimizaion} formulates the average sum-rate maximization problem for the joint design of the digital and MiLAC beamformers and develops an iterative algorithm to solve it. Section~\ref{sec:simulations} presents numerical results to evaluate the effectiveness of the proposed design. Finally, Section~\ref{sec:conclusions} concludes the paper.

\textit{Notations:}
Boldface lowercase and uppercase letters denote column vectors and matrices, respectively, while non-bold letters denote scalars. $\mathbb{R}$ and $\mathbb{C}$ denote the set of real and complex numbers. $(\cdot)^*$, $(\cdot)^T$, $(\cdot)^H$, and $(\cdot)^{-1}$ denote the conjugate, transpose, Hermitian transpose, and inverse, respectively. $\Re\{\cdot\}$, $\Im\{\cdot\}$, and $\mathbb{E}\{\cdot\}$ denote the real part, imaginary part, and expectation, respectively. $|\cdot|$, $\| \cdot \|_2$ and $\| \cdot \|_F$ refer to the absolute-value, Euclidean/spectral norm, and Frobenius norm, respectively. $\mathrm{diag}(\cdot)$, $\mathrm{blkdiag}(\cdot)$, $\mathrm{vec}(\cdot)$, and $\mathrm{Tr}(\cdot)$ denote the diagonalization, block diagonalization, vectorization, and trace operations, respectively. $\otimes$ denotes the Kronecker product. $\mathbf I_M$, $\mathbf 0_{M\times N}$, and $\mathbf 1_M$ denote the $M\times M$ identity matrix, the $M\times N$ all-zero matrix, and the $M$-dimensional all-one vector, respectively. $\mathbf A\succeq\mathbf 0$ and $\mathbf A\succ\mathbf 0$ indicate that $\mathbf A$ is positive semidefinite and positive definite, respectively. $[\mathbf a]_i$, $[\mathbf A]_{i,j}$ and $[\mathbf A]_{i:i',j:j'}$ denote an entry of $\mathbf a$, $\mathbf A$, and a submatrix of $\mathbf A$, respectively. Finally, $\jmath=\sqrt{-1}$ denotes the imaginary unit.

\section{Wideband Hybrid Digital-MiLAC System Model}\label{sec:system_model_hybrid}

\subsection{Hybrid Digital-MiLAC OFDM System Model}

\begin{figure*}
    \centering
    \includegraphics[width=0.9\textwidth]{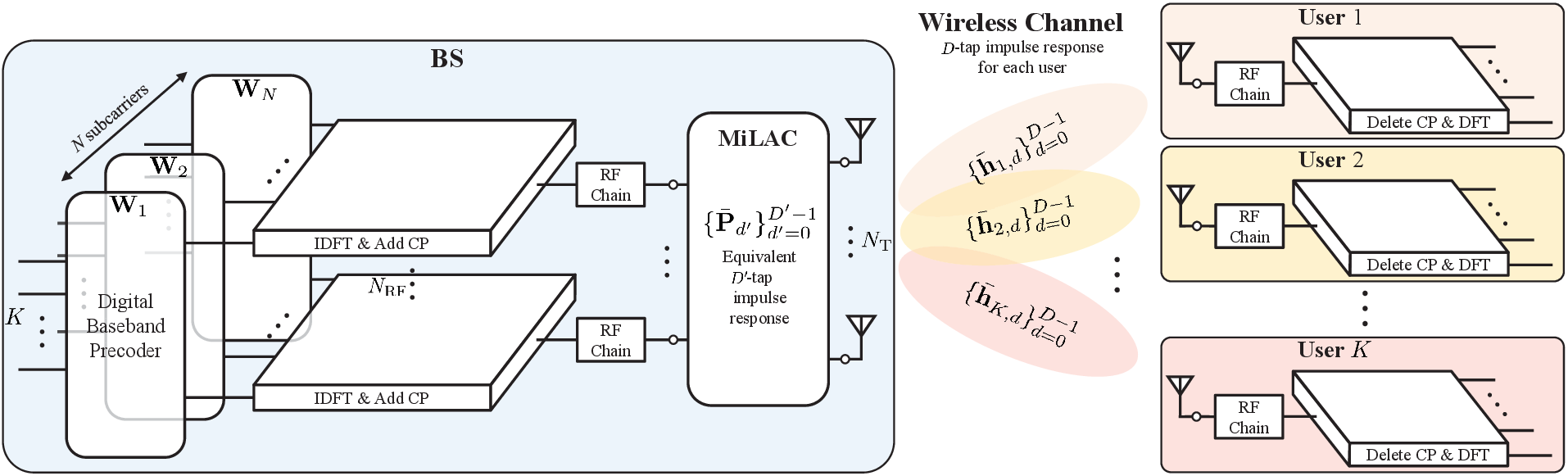}
    \caption{A hybrid digital-MiLAC-aided MU-MISO-OFDM system.}\label{fig:system_model}
\end{figure*}

We consider a wideband MU-MISO system, employing OFDM with $N$ subcarriers and a hybrid digital-MiLAC beamforming architecture. The base station (BS) is equipped with $N_\mathrm{T}$ transmit antennas and $N_\mathrm{RF}$ RF chains, serving $K$ single-antenna users scheduled over the entire bandwidth. Let $\mathcal{N}\triangleq\{1,2,\ldots,N\}$ and $\mathcal{K}\triangleq\{1,2,\ldots,K\}$ denote the index sets of subcarriers and users, respectively. Fig.~\ref{fig:system_model} illustrates the considered wideband hybrid digital-MiLAC architecture. At subcarrier $n$, the BS transmits the symbol vector $\mathbf{s}_n= [s_{1,n},s_{2,n},\ldots,s_{K,n}]^T\in\C^{K\times 1}$, where $s_{k,n}$ denotes the information symbol intended for user $k$ on subcarrier $n$, and $\mathbb{E}\{\mathbf{s}_n\mathbf{s}_n^H\}=\mathbf{I}_K,~\forall n\in\mathcal{N}$. In the frequency domain, the BS first precodes $\mathbf{s}_n$ by a digital precoder $\mathbf{W}_n = [\mathbf{w}_{1,n},\mathbf{w}_{2,n},\ldots,\mathbf{w}_{K,n}]\in\C^{N_\mathrm{RF}\times K}$, where $\mathbf{w}_{k,n}\in\C^{N_\mathrm{RF}\times 1}$ denotes the digital precoder for user $k$ on the $n$-th subcarrier. The overall digitally precoded frequency-domain signal $\mathbf{s}'\in\C^{N_\mathrm{RF}N\times 1}$ can be written as
\begin{equation}
    \mathbf{s}'=\mathbf{W}\mathbf{s},\label{eq:s'}
\end{equation}
where $\mathbf{W}=\mathrm{blkdiag}(\mathbf{W}_1,\mathbf{W}_2,\ldots,\mathbf{W}_N)$, and $\mathbf{s}=[\mathbf{s}_1^T,\mathbf{s}_2^T,\ldots,\mathbf{s}_N^T]^T$. Then $\mathbf{s}'$ is converted to the time domain through $N_\mathrm{RF}$ parallel $N$-point inverse discrete Fourier transforms (IDFTs), followed by CP insertion. Subsequently, the resulting signals are converted to analog and up-converted to passband through the RF chains before entering the MiLAC.

In the wideband system, the MiLAC exhibits a frequency-dependent response due to the frequency dependence of its reactive circuit components, e.g., tunable varactors, and the resulting multiport network response. We represent the frequency-dependent MiLAC by a $D'$-tap equivalent discrete-time baseband approximation with impulse response $\{\bar{\mathbf P}_{d'}\}_{d'=0}^{D'-1}$, where $\bar{\mathbf P}_{d'}\in\mathbb C^{N_{\mathrm T}\times N_{\mathrm{RF}}},~\forall d'\in\mathcal D'\triangleq\{0,1,\ldots,D'-1\}$, and $D'$ denotes the finite-memory length. The wireless channel between the BS and user $k$ is modeled by a $D$-tap impulse response $\{\bar{\mathbf{h}}_{k,d}\}_{d=0}^{D-1}$, with $\bar{\mathbf{h}}_{k,d}\in\C^{N_\mathrm{T}\times 1},~\forall d\in\mathcal{D}=\{0,1,\ldots,D-1\},~\forall k\in\mathcal{K}$. We assume that the MiLAC circuit configuration and the wireless channels remain constant over the CP and useful OFDM interval.

The digitally precoded time-domain signal first passes through the MiLAC and is subsequently transmitted over the wireless channel. These two linear systems form a cascade, whose equivalent impulse response is given by the convolution
\begin{equation}
    \bar{\mathbf{g}}_{k,\ell}^H
    =\sum_{d=0}^{D-1}\bar{\mathbf h}_{k,d}^H
    \bar{\mathbf P}_{\ell-d},~\ell=0,1,\ldots,D+D'-2,
    \label{eq:milac_channel_cascade_taps}
\end{equation}
where $\bar{\mathbf g}_{k,\ell}\in\mathbb C^{N_{\mathrm{RF}}\times1}$ and $\bar{\mathbf P}_j=\mathbf0$ for $j\notin\mathcal{D'}$. Within this finite-memory model, we assume that the CP covers the memory of the cascade, i.e., $N_{\mathrm{CP}}\geq D+D'-2$. Hence, after down-conversion, sampling, and CP removal, the useful received OFDM block can be represented by circular convolution of the digitally precoded time-domain signal with $\{\bar{\mathbf g}_{k,\ell}^{H}\}_{\ell=0}^{D+D'-2}$, plus additive white Gaussian noise (AWGN)~\cite{mimo_ofdm}. Applying the $N$-point DFT and using the convolution theorem yields the cascaded frequency response $\mathbf h_{k,n}^H\mathbf P_n$, where $\mathbf h_{k,n}^H=\sum_{d=0}^{D-1}\bar{\mathbf h}_{k,d}^He^{-\jmath\frac{2\pi}{N}(n-1)d}$, and $\mathbf P_n=\sum_{d'=0}^{D'-1}\bar{\mathbf P}_{d'}e^{-\jmath\frac{2\pi}{N}(n-1)d'},~\forall n\in\mathcal N$. Here, $\mathbf h_{k,n}\in\mathbb C^{N_{\mathrm T}\times1}$ denotes the frequency-domain channel at the $n$-th subcarrier for user $k$, and $\mathbf P_n\in\mathbb C^{N_{\mathrm T}\times N_{\mathrm{RF}}}$ denotes the MiLAC response at subcarrier $n$. Consequently, the frequency-domain received signal at subcarrier $n$ for user $k$ is
\begin{equation}
    r_{k,n}=\mathbf h_{k,n}^H\mathbf P_n\mathbf W_n\mathbf s_n+z_{k,n},~\forall n\in\mathcal N,~\forall k\in\mathcal K,
    \label{eq:hybrid_milac_received_signal}
\end{equation}
where $z_{k,n}\sim\mathcal{CN}(0,\sigma^2)$ denotes AWGN.

\subsection{Frequency-Dependent MiLAC Modeling}
A MiLAC can be modeled as a reconfigurable microwave network \cite{nerini_milac_partI,nerini_milac_partII}. In the considered hybrid digital-MiLAC beamforming architecture, the MiLAC is represented as an $(N_{\mathrm{RF}}+N_{\mathrm{T}})$-port network, where the first $N_{\mathrm{RF}}$ ports are connected to the RF chains and the remaining $N_{\mathrm{T}}$ ports are connected to the transmit antennas. In this paper, we consider the fully-connected MiLAC architecture studied in \cite{nerini_milac_partI,nerini_milac_partII,nerini_milac_capacity,zheyu_milac_beamforming,tianyu_milac_beamforming}, where each port is connected to the ground through a tunable component and is also interconnected to every other port through tunable components. Define $N_\mathrm{tot}\triangleq N_{\mathrm{RF}}+N_{\mathrm{T}}$ and $\mathcal N_{\mathrm{tot}}\triangleq\{1,2,\ldots,N_{\mathrm{tot}}\}$. An example of a four-port fully-connected MiLAC is shown in Fig.~\ref{fig:MiLAC and circuit}.

To characterize the wideband MiLAC, we evaluate the response of the microwave network at each subcarrier frequency $f_n$, with angular
frequency $\omega_n=2\pi f_n$. The corresponding admittance and scattering matrices are denoted by $\mathbf Y_n\in\mathbb C^{N_{\mathrm{tot}}\times N_{\mathrm{tot}}}$ and $\bm\Phi_n\in\mathbb C^{N_{\mathrm{tot}}\times N_{\mathrm{tot}}}$, respectively. At subcarrier $n$, the frequency-domain MiLAC input-output response $\mathbf P_n\in\mathbb C^{N_{\mathrm T}\times N_{\mathrm{RF}}}$ is given by \cite{nerini_milac_partI,nerini_milac_partII}
\begin{equation}
    \mathbf P_n
    =\left[(\mathbf I_{N_{\mathrm{tot}}}+Z_0\mathbf Y_n)^{-1}\right]_{N_{\mathrm{RF}}+1:N_{\mathrm{tot}},1:N_{\mathrm{RF}}},
    \label{eq:P_to_Y}
\end{equation}
where $Z_0=50\Omega$ denotes the reference impedance. The scattering matrix satisfies \cite{pozar2009microwave}
\begin{equation}
    \bm\Phi_n
    =(\mathbf I_{N_{\mathrm{tot}}}+Z_0\mathbf Y_n)^{-1}
    (\mathbf I_{N_{\mathrm{tot}}}-Z_0\mathbf Y_n).
    \label{eq:Y_to_Phi}
\end{equation}
Rewriting \eqref{eq:Y_to_Phi} as $\bm\Phi_n=2(\mathbf I_{N_{\mathrm{tot}}}+Z_0\mathbf Y_n)^{-1}
-\mathbf I_{N_{\mathrm{tot}}}$ and extracting the input-to-antenna block yields \cite{nerini_milac_capacity,nerini_milac_stem}
\begin{equation}
    \mathbf P_n
    =\frac12[\bm\Phi_n]_{N_{\mathrm{RF}}+1:N_{\mathrm{tot}},1:N_{\mathrm{RF}}}.
    \label{eq:P_to_Phi}
\end{equation}
In this paper, we consider a lossless and reciprocal MiLAC with a frequency-dependent circuit response. We next present the details.

\subsubsection{Lossless and Reciprocal MiLAC}Let $Y_{m,m',n}$ denote the tunable admittance component connecting port $m$ to port $m'$ at subcarrier $n,~\forall m\neq m'$, and let $Y_{m,m,n}$ denote the tunable admittance component connecting port $m$ to the ground. The reciprocal constraint arises from the fact that the tunable components employed in the microwave network, e.g., varactors, are typically reciprocal devices, rendering $Y_{m,m',n}=Y_{m',m,n}$ for $m'\neq m$. The admittance matrix is constructed as \cite{pozar2009microwave}
\begin{equation}
    [\mathbf{Y}_n]_{m,m'} = \begin{cases}
            -Y_{m,m',n}, &\text{if}~~m \neq m'; \\
            \sum_{q=1}^{N_\mathrm{tot}} Y_{m,q,n}, &\text{if}~~m = m',
        \end{cases}\label{eq:Y_component_mapping}
\end{equation}
for all $m,m'\in\mathcal N_{\mathrm{tot}}$ and $n\in\mathcal N$, which mathematically yields a symmetric admittance matrix $\Y_n$. Moreover, all admittance components in a lossless reciprocal MiLAC are purely imaginary, i.e., $\Re\{Y_{m,m',n}\}=0,~\forall m,m'\in\mathcal{N}_\mathrm{tot},~\forall n\in\mathcal N$. Hence, each admittance matrix can be written as $\mathbf Y_n=\jmath\mathbf{B}_n$, where $\mathbf{B}_n\in\mathbb{R}^{N_{\mathrm{tot}}\times N_{\mathrm{tot}}}$ is the susceptance matrix of MiLAC at subcarrier $n$. The susceptance component at subcarrier $n$ is denoted by $B_{m,m',n}\triangleq\Im\{Y_{m,m',n}\}$,~$\forall m,m'\in\mathcal N_{\mathrm{tot}}$,~$\forall n\in\mathcal N$.

\subsubsection{Frequency-Dependent Modeling of Tunable Admittance}

\begin{figure}
    \centering
    \includegraphics[width=0.45\textwidth]{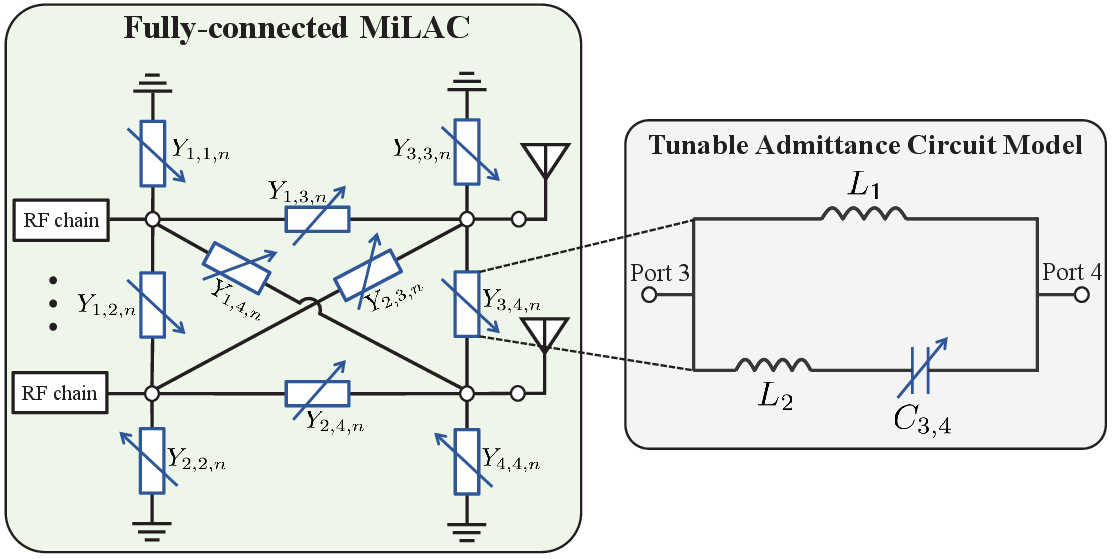}
    \caption{A fully-connected four-port MiLAC with two input and two output ports, and the frequency-dependent circuit model for each tunable admittance.}\label{fig:MiLAC and circuit}
\end{figure}

To specify the frequency dependence of these components, we adopt a lumped circuit model consisting of two fixed inductors $L_1$ and $L_2$ and one tunable capacitor $C_{m,m'}$ \cite{koziel2013surrogate,wideband_circuit_hongyu}, as illustrated in Fig.~\ref{fig:MiLAC and circuit}. At angular frequency $\omega=2\pi f$, its admittance is
\begin{equation}
    Y_{m,m'}(C_{m,m'},\omega)= \frac{1}{\jmath\omega L_1} + \frac{1}{\jmath \omega L_2 + \frac{1}{\jmath \omega C_{m,m'}}},
    \label{eq:admittance_C_f}
\end{equation}
$\forall m,m'\in\mathcal{N}_\mathrm{tot}$, and thus $Y_{m,m',n}=Y_{m,m'}(C_{m,m'},\omega_n)$. In this circuit model, the inductances are fixed and identical for all tunable admittance components, while the capacitance is reconfigurable. Once $C_{m,m'}$ is selected, \eqref{eq:admittance_C_f} determines the response of that component at every subcarrier.

Let $f_{\mathrm c}$ be the center frequency and $\omega_{\mathrm c}=2\pi f_{\mathrm c}$. Define $B_{\mathrm c,m,m'}=\Im\{Y_{m,m'}(C_{m,m'},\omega_{\mathrm c})\}$ as the center-frequency susceptance. Based on the findings in \cite{wideband_circuit_hongyu}, the capacitance $C_{m,m'}$ can be expressed in terms of $B_{\mathrm{c},m,m'}$ as
\begin{equation}
    C_{m,m'} = \left(\omega_\mathrm{c}^2 L_2 + \frac{\omega_\mathrm{c}}{B_{\mathrm{c},m,m'} + \frac{1}{\omega_\mathrm{c} L_1}}\right)^{-1}.
    \label{eq:C_to_Bc}
\end{equation}
Thus, $B_{\mathrm c,m,m'}$ can be used as the tunable circuit variable. The practical capacitance range of each component can be equivalently mapped to a feasible range of each $B_{\mathrm{c},m,m'}$. Specifically, $B_{\mathrm{c},m,m'}\in[B_{\mathrm{min}},B_{\mathrm{max}}],~\forall m,m'\in\mathcal{N}_\mathrm{tot}$, where $B_{\mathrm{min}}$ and $B_{\mathrm{max}}$ are determined by the minimum and maximum feasible capacitance values, respectively.

Following \cite{wideband_circuit_hongyu}, we use an LC model based on the SMV1231-079LF varactor, with $L_1=2.5$ nH, $L_2=0.7$ nH, and
$C_{m,m'}\in[0.2,3]$ pF. In this paper, we consider a wideband system having center frequency $f_\mathrm{c}=2.4$ GHz and bandwidth $B=100$ MHz, covering $2.35$--$2.45$ GHz. In this operating band, the frequency dependence of each susceptance component is approximated by the linear fitting model in \cite{wideband_circuit_hongyu}
\begin{equation}
    B_{m,m',n}= F_1(\omega_n)B_{\mathrm{c},m,m'} + F_2(\omega_n),\label{eq:B_linear_model}
\end{equation}
$\forall m,m'\in\mathcal{N}_\mathrm{tot},~\forall n\in\mathcal{N}$, where $F_1(\omega_n) = \alpha_1 \omega_n + \beta_1$ and $F_2(\omega_n) = \alpha_2 \omega_n + \beta_2$.\footnote{The parameters $\alpha_1,\beta_1,\alpha_2,\beta_2$ are determined by circuit parameters $L_1$, $L_2$, and $\omega_\mathrm{c}$. In this work, we use $\alpha_1 = 2.0046\times10^{-10}$, $\beta_1 = -1.9968$, $\alpha_2 = 6.2775\times10^{-12}$, and $\beta_2 = -0.0942$ as in \cite{wideband_circuit_hongyu}. More details about the linear model \eqref{eq:B_linear_model} can be found in \cite{wideband_circuit_hongyu}.} With this model, the center-frequency susceptance $B_{\mathrm{c},m,m'}$ serves as the tunable variable of each component, while its susceptance values over all OFDM subcarriers are determined by \eqref{eq:B_linear_model}. Therefore, the susceptance values of the same component at different subcarriers are coupled through one common tunable variable, which reflects the frequency coupling induced by the common tunable capacitor. 

\section{Wideband MiLAC-aided Beamforming Flexibility and Limitations}\label{sec:wideband_milac_flexibility}
In this section, we explain how one circuit in wideband MiLAC-aided beamforming jointly determines the responses across subcarriers, and use a MiLAC with one input port and two output ports as an example to illustrate its impact.

\subsection{Frequency Coupling of the Wideband MiLAC Beamforming}
MiLAC-aided beamforming is a special case of the hybrid digital-MiLAC architecture, where $K$ data streams are directly fed into $K$ MiLAC input ports without digital precoding. Accordingly, throughout this section, we specialize the system model in Section~\ref{sec:system_model_hybrid} to
$N_{\mathrm{RF}}=K$ and $N_{\mathrm{tot}}=K+N_{\mathrm T}$. Let $\mathbf P_n^{\mathrm M}\in\mathbb C^{N_{\mathrm T}\times K}$ denote the MiLAC-aided beamforming response at subcarrier $n$. The received signal of user $k$ is
\begin{equation}
    r_{k,n}^\mathrm{M}=\mathbf{h}_{k,n}^H\mathbf{P}^\mathrm{M}_n\mathrm{diag}(\mathbf{p}^\mathrm{D}_n)^{\frac{1}{2}}\mathbf{s}_n+z_{k,n},~\forall n\in\mathcal{N},~\forall k\in\mathcal{K},
\end{equation}
where $\mathbf{p}^\mathrm{D}_n=[p^\mathrm{D}_{1,n},p^\mathrm{D}_{2,n},\ldots,p^\mathrm{D}_{K,n}]\in\mathbb{R}^{K\times 1}$ with $p^\mathrm{D}_{k,n}\geq 0$ denoting the digital power allocated to symbol $s_{k,n}$. From \eqref{eq:P_to_Y}--\eqref{eq:P_to_Phi}, we have the MiLAC input-output relations $\mathbf{P}_n^\mathrm{M} = \left[(\mathbf{I}_{N_{\mathrm{tot}}}+\jmath Z_0\mathbf{B}_n)^{-1}\right]_{K+1:N_{\mathrm{tot}},1:K}$, and $\mathbf{P}_n^\mathrm{M}=\frac{1}{2}[\bm\Phi_n]_{K+1:N_{\mathrm{tot}},1:K}$, respectively. 

Although the MiLAC response varies across subcarriers, its component settings are common to all subcarriers. Specifically, each center-frequency susceptance $B_{\mathrm c,m,m'}$ determines the corresponding susceptances $\{B_{m,m',n}\}_{n=1}^{N}$ through \eqref{eq:B_linear_model}. Consequently, the susceptance matrices $\{\mathbf B_n\}_{n=1}^{N}$ and the beamforming responses $\{\mathbf P_n^{\mathrm M}\}_{n=1}^{N}$ are jointly determined by one physical MiLAC circuit configuration, i.e., one fixed choice of the center-frequency susceptances. Frequency dependence therefore provides different beamforming responses across subcarriers, but does not provide separate circuit controls for each subcarrier.

A general analytical characterization of the beamforming magnitudes and directions is challenging, since the MiLAC response depends nonlinearly on the bounded susceptance components through the matrix inverse in \eqref{eq:P_to_Y}. Moreover, knowing the shared circuit configuration does not directly reveal how the center-frequency susceptances limit the beamforming magnitudes and directions achievable across subcarriers. The following subsection illustrates these effects by evaluating discretized circuit configurations for a MiLAC with one input port and two output ports.

\subsection{Visualization of Frequency-Dependent MiLAC Beamforming Flexibility}
\label{subsec:milac_visualization}

\begin{figure*}[!t]
    \centering
    \subfloat[Beamforming gain ratio]{
        \includegraphics[width=0.32\textwidth]{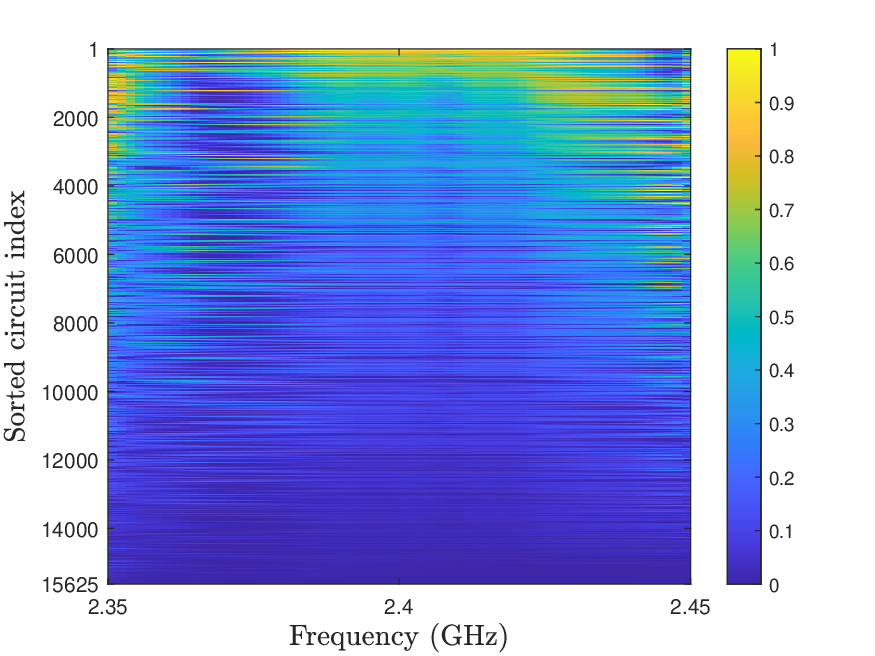}
        \label{fig:milac_gain_heatmap}
    }
    \subfloat[Magnitude efficiency]{
        \includegraphics[width=0.32\textwidth]{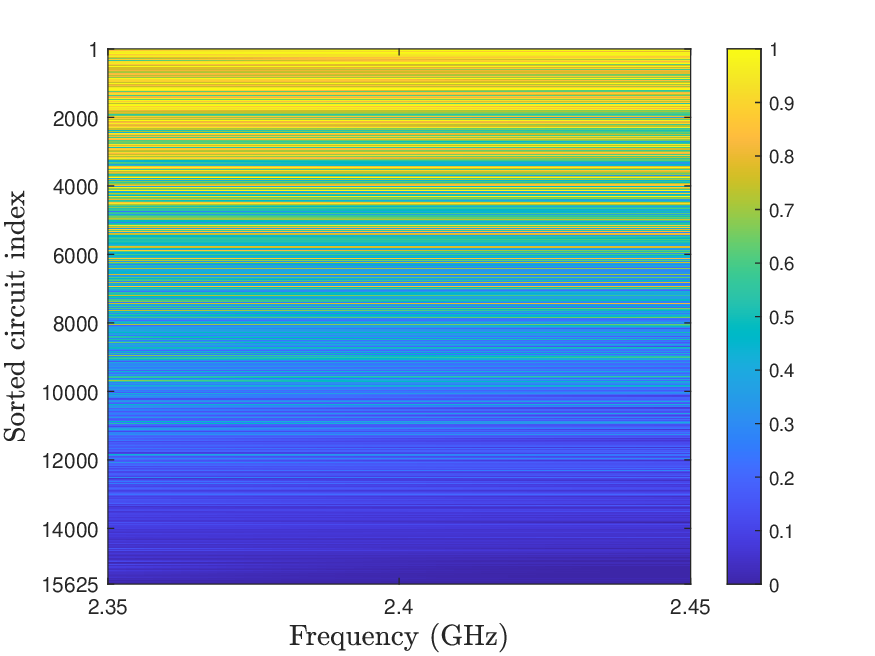}
        \label{fig:milac_mag_eff_heatmap}
    }
    \subfloat[Directional alignment]{
        \includegraphics[width=0.32\textwidth]{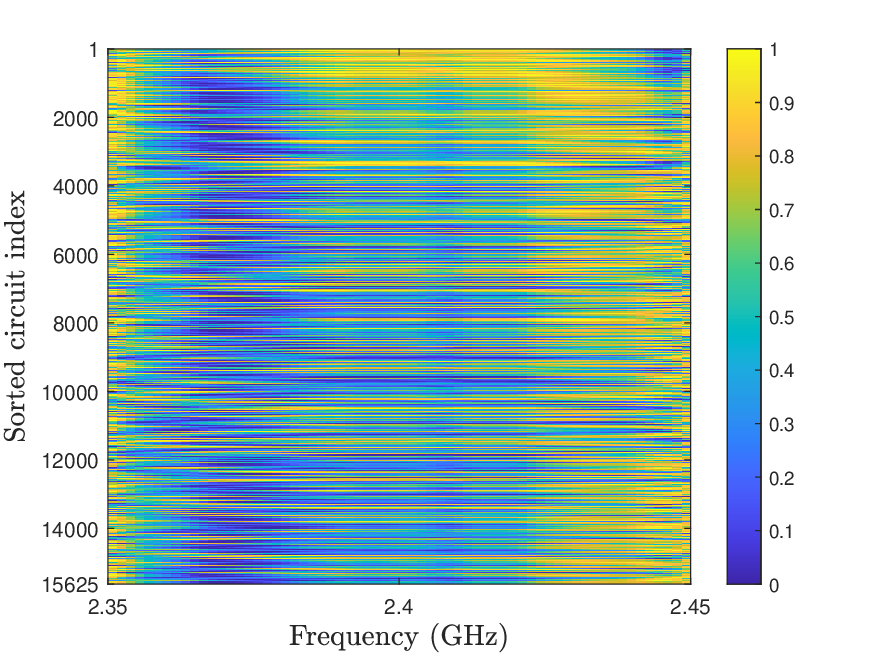}
        \label{fig:milac_alignment_heatmap}
    }
    \hspace{-2mm}
    \subfloat[Gain ratio of the selected circuits]{
        \includegraphics[width=0.32\textwidth]{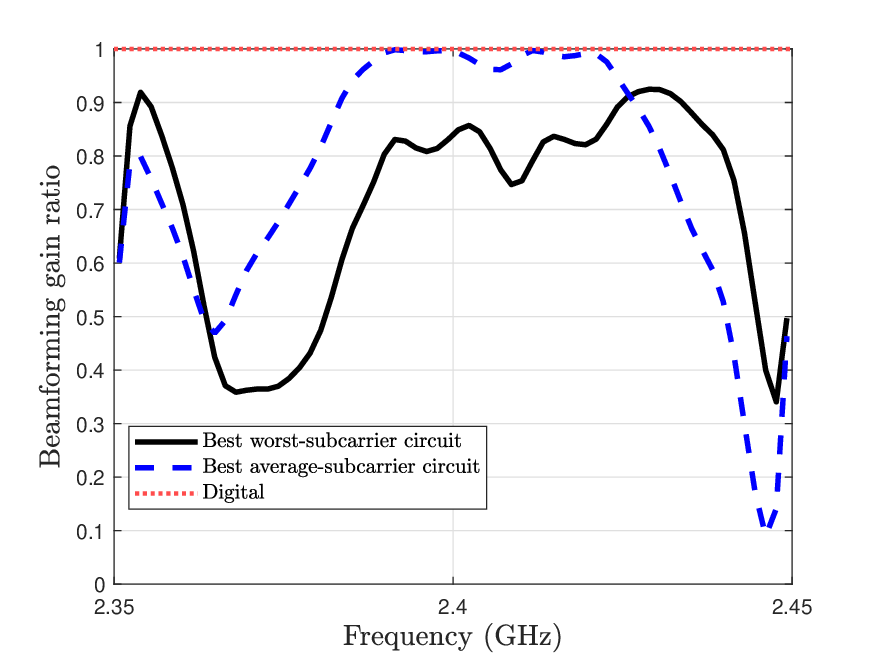}
        \label{fig:milac_representative_curves}
    }
    \subfloat[Gain decomposition of $l_{\mathrm{avg}}$]{
        \includegraphics[width=0.32\textwidth]{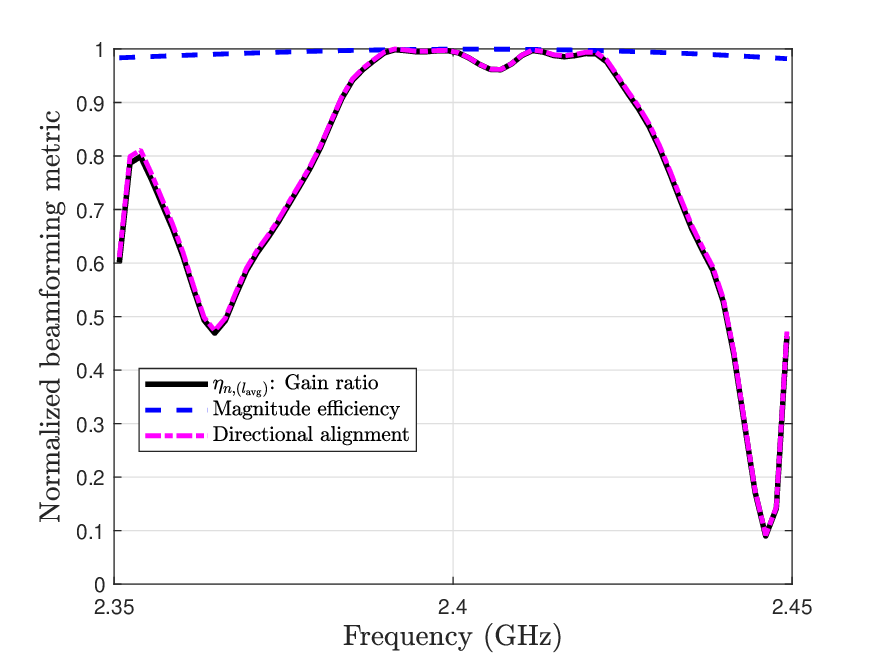}
        \label{fig:milac_gain_decomposition}
    }
    \subfloat[Normalized complex coefficient of
    $l_{\mathrm{avg}}$]{
        \includegraphics[width=0.32\textwidth]{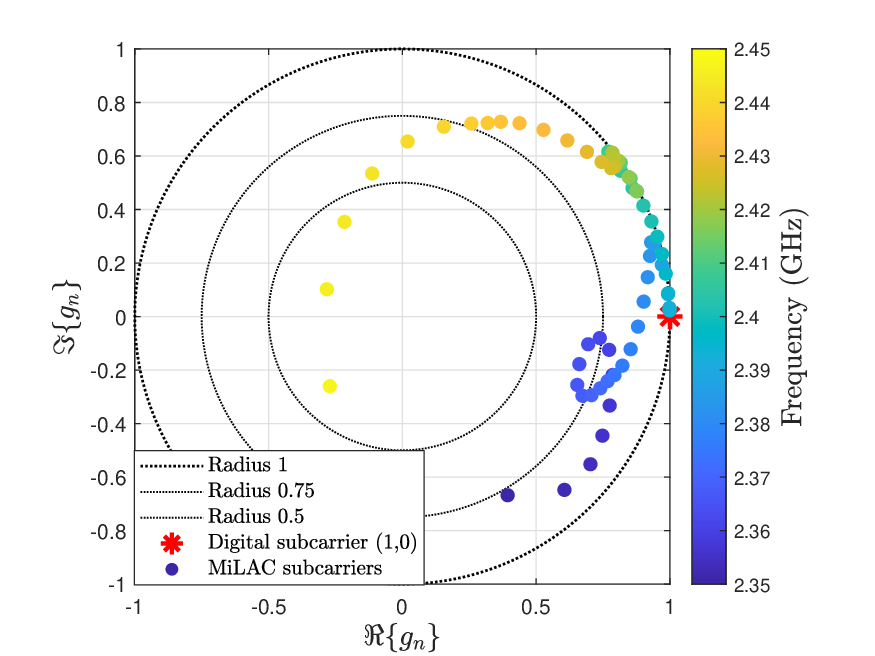}
        \label{fig:milac_complex_gain}
    }
    \caption{
    Visualization of wideband MiLAC beamforming flexibility for $K=1$, $N_{\mathrm T}=2$, and $N=64$, over a $100$ MHz bandwidth centered at $2.4$ GHz. (a)--(c) Gain ratio, magnitude efficiency, and directional alignment of the $5^6$ sampled circuit configurations. The circuits are sorted in descending order of their average gain ratio $\frac{1}{N}\sum_{n=1}^{N}\eta_{n,(l)}$, and the same ordering is used in (a)--(c). (d) Gain ratios of $l_{\mathrm{mm}}$ and $l_{\mathrm{avg}}$. (e) Gain ratio and its two factors for $l_{\mathrm{avg}}$. (f) Normalized complex coefficient $\{g_n\}_{n=1}^N$ for $l_{\mathrm{avg}}$.
    }
    \label{fig:milac_only_visualization}
\end{figure*}

We consider a single-user system with $N_{\mathrm T}=2$ and $N=64$. The fully-connected lossless and reciprocal MiLAC has one input port, two output ports, and six tunable components, whose susceptances at $n$-th subcarrier are $B_{1,1,n},B_{1,2,n},B_{1,3,n},B_{2,2,n},B_{2,3,n}$, and $B_{3,3,n}$. These tunable susceptances construct the MiLAC susceptance matrix $\mathbf{B}_n\in\mathbb{R}^{3\times 3}$ as
\begin{equation}
    [\mathbf{B}_n]_{m,m'} = \begin{cases}
            -B_{m,m',n}, &\text{if}~~m \neq m'; \\
            \sum_{q=1}^{3} B_{m,q,n}, &\text{if}~~m = m',
        \end{cases}
\end{equation}
for $m,m'=1,2,3,~\forall n\in\mathcal{N}$. Each susceptance matrix can be converted to the corresponding scattering matrix $\bm\Phi_n\in\C^{3\times 3}$ through \eqref{eq:Y_to_Phi}. We sample each center-frequency susceptance using five values in $[B_{\mathrm{min}},B_{\mathrm{max}}]$, giving $5^6=15625$ circuit configurations, indexed by $l\in\mathcal L_\mathrm{c}\triangleq\{1,2,\ldots,5^6\}$. For configuration $l$, let $\mathbf p_{n,(l)}^{\mathrm M}=\frac12[\bm\Phi_{n,(l)}]_{2:3,1}\in\mathbb C^{2\times1}$ denote the MiLAC-aided beamforming vector at subcarrier $n$ of $l$-th circuit. Each configuration fixes all six center-frequency susceptances and thereby determines the entire response sequence $\{\mathbf p_{n,(l)}^{\mathrm M}\}_{n=1}^{N}$. The wideband channel is generated from $16$ independent taps, denoted by $\bar{\mathbf h}_d\sim\mathcal{CN}(\mathbf0,p_d\mathbf I_2)$, where normalized power delay profile (PDP) satisfies $p_d=\frac{e^{-d/\epsilon}}{\sum_{i=0}^{15}e^{-i/\epsilon}}$, $d=0,1,\ldots,15$, with $\epsilon=1.5$ being the PDP decay factor.

To benchmark the MiLAC-aided beamforming flexibility, we compare it with digital matched beamforming. Following the discussion in \cite{nerini_milac_capacity}, when transmit antennas are perfectly matched to the source-generator series impedance $Z_0$, the digital matched beamformer is
$\mathbf w_n^{\mathrm D}
=\frac{\mathbf{h}_n}{\|\mathbf{h}_n\|_2}$, where $\mathbf{h}_n\in\C^{2\times 1}$ is the channel at subcarrier $n$. Its received beamforming gain is $G_n^{\mathrm D}=|\frac{1}{2}\mathbf h_n^H\mathbf w_n^{\mathrm D}|^2=\frac14\|\mathbf h_n\|_2^2$.\footnote{The $1/2$ factor stems from the fact that for transmit antennas perfectly matching the series impedance $Z_0$ of the source signal, voltage division makes the amplitude of the transmit signal equal to half that of the source signal~\cite{nerini_milac_capacity}.} The beamforming gain ratio, defined as the MiLAC beamforming gain normalized by the digital matched-beamforming gain, is
\begin{equation}
    \eta_{n,{(l)}}
    =
    \frac{
    \left|\mathbf{h}_n^H\mathbf{p}_{n,(l)}^{\mathrm{M}}\right|^2
    }{
    G_n^{\mathrm D}
    }=
    \underbrace{
    \left(
    \frac{\|\mathbf{p}_{n,(l)}^{\mathrm{M}}\|_2}{\frac{1}{2}}
    \right)^2
    }_{\chi^\mathrm{mag}_{n,(l)}}
    \underbrace{
    \frac{
    \left|\mathbf{h}_n^H\mathbf{p}_{n,(l)}^{\mathrm{M}}\right|^2
    }{
    \|\mathbf{h}_n\|_2^2\|\mathbf{p}_{n,(l)}^{\mathrm{M}}\|_2^2
    }
    }_{\chi^\mathrm{dir}_{n,(l)}}.
    \label{eq:normalized_milac_gain}
\end{equation}
Here, $\chi^\mathrm{mag}_{n,(l)}$ is the magnitude efficiency and $\chi^\mathrm{dir}_{n,(l)}$ is the directional alignment. The unitarity of the scattering matrix gives $\chi^\mathrm{mag}_{n,(l)}=1-|[\bm\Phi_{n,(l)}]_{1,1}|^2\leq1$. Moreover, $0\leq\chi^\mathrm{dir}_{n,(l)}\leq1$ by the Cauchy-Schwarz inequality. $\eta_{n,(l)}=1$ requires both maximum magnitude and perfect directional alignment. To compare the MiLAC beamforming obtained with different design criteria, we select the circuit maximizing the worst-subcarrier gain ratio as
\begin{equation}
    l_{\mathrm{mm}}
    =\arg\max_{l\in\mathcal L_\mathrm{c}}
    \min_{n\in\mathcal N}\eta_{n,(l)},
    \label{eq:best_worst_circuit}
\end{equation}
and the circuit maximizing average-subcarrier gain ratio as
\begin{equation}
    l_{\mathrm{avg}}
    =\arg\max_{l\in\mathcal L_\mathrm{c}}
    \frac{1}{N}\sum_{n=1}^{N}\eta_{n,(l)}.
    \label{eq:best_avg_circuit}
\end{equation}
To visualize the response of the best average-subcarrier circuit in the complex plane, we define its normalized complex coefficient at subcarrier $n$ as $g_n\triangleq\frac{\mathbf h_n^H\mathbf p_{n,(l_{\mathrm{avg}})}^{\mathrm M}}{\mathbf h_n^H\mathbf w_n^{\mathrm D}}=\frac{\mathbf h_n^H\mathbf p_{n,(l_{\mathrm{avg}})}^{\mathrm M}}{\frac12\|\mathbf h_n\|_2}$. 

Fig.~\ref{fig:milac_only_visualization} shows that the sampled MiLAC circuits can achieve high gain ratios at some subcarriers but exhibit substantial gain reductions at others. For the $l_{\mathrm{avg}}$ circuit, the magnitude efficiency remains close to one, while the gain ratio largely follows the directional alignment, indicating that directional mismatch is the main source of gain reduction. Although $l_{\mathrm{mm}}$ improves the minimum gain, it still cannot maintain a high gain ratio over the entire band. This behavior is also reflected in Fig.~\ref{fig:milac_complex_gain}, where the normalized responses vary considerably across subcarriers. These results illustrate that, although the MiLAC beamformer varies with frequency, one MiLAC circuit configuration cannot independently adapt its beamforming direction at each subcarrier. This motivates hybrid digital-MiLAC beamforming, where subcarrier-dependent digital precoders provide additional flexibility to compensate for the frequency-coupled MiLAC responses.

\section{Joint Wideband Hybrid Digital-MiLAC Beamforming Optimization}\label{sec:optimizaion}
In this section, we first discuss the design flexibility of wideband hybrid digital-MiLAC beamforming. We then formulate the average sum-rate maximization problem and develop a BCD-ADMM algorithm to jointly optimize the digital and MiLAC beamformers across all subcarriers.

\subsection{Design Degrees of Freedom of Wideband Hybrid Digital-MiLAC Beamforming}
\label{subsec:rf_chain_design_flexibility}

For narrowband MU-MISO systems, it was established
in~\cite{zheyu_milac_beamforming} that hybrid digital-MiLAC beamforming can realize an arbitrary digital beamforming matrix with $N_{\mathrm{RF}}=K$. Extending this characterization to wideband OFDM systems is nontrivial, since one MiLAC circuit configuration determines the beamforming responses at all subcarriers. To gain insight into this distinction, we compare the numbers of real design variables in digital and hybrid digital-MiLAC beamforming.

Let $\{\mathbf W_n^{\mathrm D}\}_{n=1}^{N}$ denote an unrestricted collection of digital beamformers, where $\mathbf W_n^{\mathrm D}\in\mathbb C^{N_{\mathrm T}\times K}$ is the precoder at the $n$-th subcarrier. The digital beamformers contain $2NN_{\mathrm T}K$ real coefficients. For hybrid digital-MiLAC beamforming, the digital precoders $\{\mathbf W_n\}_{n=1}^{N}$ contain $2NN_{\mathrm{RF}}K$ real coefficients. For fully-connected lossless reciprocal MiLAC, $\mathbf{B}_n=\mathbf{B}_n^T,~\forall n\in\mathcal{N}$, and each $\mathbf B_n$ is determined by the real susceptance components $B_{m,m',n}$, $1\leq m\leq m'\leq N_{\mathrm{tot}}$ through~\eqref{eq:Y_component_mapping}, where $N_{\mathrm{tot}}=N_{\mathrm T}+N_{\mathrm{RF}}$. Furthermore, \eqref{eq:B_linear_model} determines the susceptances $\{B_{m,m',n}\}_{n=1}^{N}$ of each component from its center-frequency value $B_{\mathrm c,m,m'}$. The MiLAC circuit therefore has $\tilde N_{\mathrm{tot}}\triangleq N_{\mathrm{tot}}(N_{\mathrm{tot}}+1)/2$ real control variables shared across all subcarriers. We collect these center-frequency susceptances in $\mathbf b_{\mathrm c}\in\mathbb R^{\tilde N_{\mathrm{tot}}\times1}$, ordered as $\mathbf{b}_\mathrm{c}^T=[B_{\mathrm{c},1,1},\ldots,B_{\mathrm{c},1,N_\mathrm{tot}},B_{\mathrm{c},2,2},\ldots,B_{\mathrm{c},2,N_\mathrm{tot}},\ldots,B_{\mathrm{c},N_\mathrm{tot},N_\mathrm{tot}}]$. 

Together with the digital precoders, hybrid digital-MiLAC architecture is thus parameterized by $\tilde{N}_{\mathrm{tot}}+2NN_{\mathrm{RF}}K$ real variables. For fixed $N_{\mathrm T}$ and $N_{\mathrm{RF}}$, the number of MiLAC circuit controls does not increase with $N$. Hence, reducing $N_{\mathrm{RF}}$ removes digital design variables at every subcarrier, while the independent MiLAC circuit controls remain shared across the entire bandwidth. For example, with $N_{\mathrm T}=64$, $K=4$, and $N=64$, digital beamforming contains $32768$ real coefficients, whereas hybrid digital-MiLAC beamforming with $N_{\mathrm{RF}}=K=4$ is parameterized by only $4394$ real variables. This comparison motivates jointly exploiting the subcarrier-specific digital precoders and the shared MiLAC circuit controls to achieve high wideband beamforming performance with reduced RF chains. Accordingly, in Section~\ref{subsec:problem_formulate}, we jointly optimize the digital and MiLAC beamformers across all subcarriers for a general number of RF chains to assess the performance of the hybrid architecture in OFDM systems.

\subsection{Problem Formulation and WMMSE Reformulation}\label{subsec:problem_formulate}
We aim to maximize the average sum-rate with hybrid digital-MiLAC beamformer subject to the constraints of lossless reciprocal MiLAC and a total radiated-power budget in the MU-MISO-OFDM system. The problem is formulated as
\begin{subequations}\label{eq:max_sum_rate}
    \begin{align}                    \max_{\substack{\{\mathbf{P}_n\},\{\mathbf{W}_n\}}}~& \frac{1}{N}\sum_{n=1}^N\sum_{k=1}^K  \log_2 \left(1+\frac{|\mathbf{h}_{k,n}^H\mathbf{P}_n \mathbf{w}_{k,n}|^2}{\sum\limits_{j\neq k}|\mathbf{h}_{k,n}^H\mathbf{P}_n \mathbf{w}_{j,n}|^2+\sigma^2}\right)\label{eq:max_sum_rate_a}\\
    \mathrm{s.t.}~~~~~&\mathbf{P}_n = \left[(\mathbf{I}_{N_\mathrm{tot}}+\jmath Z_0\mathbf{B}_n)^{-1}\right]_{N_{\mathrm{RF}}+1:N_\mathrm{tot},1:N_{\mathrm{RF}}},\label{eq:max_sum_rate_b} \\ 
    &\mathbf{B}_n = \mathbf{B}_n^T,\label{eq:max_sum_rate_c} \\ 
    &[\mathbf{B}_n]_{m,m'} = \begin{cases}
            -B_{m,m',n}, &m \neq m' \\
            \sum_{q=1}^{N_\mathrm{tot}} B_{m,q,n}, & m = m'
        \end{cases},\label{eq:max_sum_rate_d} \\ 
    &B_{m,m',n} = F_1(\omega_n)B_{\mathrm{c},m,m'} + F_2(\omega_n),\notag\\
    &\hspace{2.3cm}\forall m,m'\in\mathcal N_{\mathrm{tot}},~\forall n\in\mathcal{N},\label{eq:max_sum_rate_e} \\ 
    &B_{\mathrm{c},m,m'} \in [B_\mathrm{min},B_\mathrm{max}],~\forall m,m'\in\mathcal N_{\mathrm{tot}},\label{eq:max_sum_rate_f}\\
    &\sum_{n=1}^N \mathrm{Tr}(\mathbf{W}_n^H\mathbf{P}_n^H\mathbf{P}_n\mathbf{W}_n )\leq P_\mathrm{T},\label{eq:max_sum_rate_g}
    \end{align}
\end{subequations}
where $\omega_n = 2\pi f_n$ and $P_\mathrm{T}$ denotes the total radiated-power budget across all subcarriers at the BS. Problem \eqref{eq:max_sum_rate} is highly challenging to solve for the following reasons. First, $\{\mathbf{P}_n\}$ and $\{\mathbf{W}_n\}$ are coupled in \eqref{eq:max_sum_rate_a} and \eqref{eq:max_sum_rate_g}. Second, the MiLAC beamformers are nonlinear functions of the susceptance matrices in \eqref{eq:max_sum_rate_b}, which involves a matrix inverse. Third, the MiLAC beamformers across subcarriers are coupled through the common center-frequency susceptances according to the frequency-dependent circuit model in \eqref{eq:max_sum_rate_e}. Finally, the large numbers of transmit antennas $N_{\mathrm{T}}$ and subcarriers $N$ further increase the computational complexity of the optimization, necessitating an efficient solution method. To address these challenges, we first reformulate the sum-rate maximization form into an equivalent WMMSE form and then iteratively design the MiLAC beamformers and digital beamformers.

Denote $u_{k,n}\in\C$ as the receive equalizer for user $k$ on subcarrier $n$. The estimated symbol of user $k$ on subcarrier $n$ is given by $\hat{s}_{k,n}=u_{k,n}^*r_{k,n}$. Then, the mean-square error (MSE) between symbol $s_{k,n}$ and estimated symbol $\hat{s}_{k,n}$, i.e., $E_{k,n}\triangleq \mathbb{E}\{|\hat{s}_{k,n}-s_{k,n}|^2\}$, can be expressed as 
\begin{align}
    E_{k,n}(u_{k,n},\mathbf{P}_n,\mathbf{W}_{n})
    &=|1-u_{k,n}^*\mathbf{h}_{k,n}^H\mathbf{P}_n\mathbf{w}_{k,n}|^2 \notag\\&+ |u_{k,n}|^2\left(\sum_{j\neq k}|\mathbf{h}_{k,n}^H\mathbf{P}_n\mathbf{w}_{j,n}|^2+\sigma^2\right).
\end{align}
By introducing weight auxiliary variables $\{\bm\nu_n\}_{n=1}^N$, and applying the transformation in \cite{WMMSE}, we can equivalently reformulate problem \eqref{eq:max_sum_rate} into the following WMMSE form:
\begin{subequations}\label{eq:max_sum_rate_wmmse}
    \begin{align}                    \min_{\substack{\{\mathbf{P}_n\},\{\mathbf{W}_n\} \\ \{\mathbf{u}_n\},\{\bm\nu_n\}}}~& \frac{1}{N}\sum_{n=1}^N\sum_{k=1}^K  \left(\frac{1}{\ln 2}\nu_{k,n}E_{k,n}-\log_2(\nu_{k,n})\right)\label{eq:max_sum_rate_wmmse_a}\\
    \mathrm{s.t.}~~~~~&\text{\eqref{eq:max_sum_rate_b}} -\text{\eqref{eq:max_sum_rate_g}},\label{eq:max_sum_rate_wmmse_b}
    \end{align}
\end{subequations}
where $\mathbf{u}_n=[u_{1,n},u_{2,n},\ldots,u_{K,n}]\in\C^{K\times 1}$ and $\bm\nu_n=[\nu_{1,n},\nu_{2,n},\ldots,\nu_{K,n}]\in\mathbb{R}^{K\times 1}$ with $\nu_{k,n}>0,~\forall n\in\mathcal{N}$. Although problem \eqref{eq:max_sum_rate_wmmse} remains nonconvex, the WMMSE objective function is convex with respect to each variable block when the other variables are fixed. To solve \eqref{eq:max_sum_rate_wmmse}, we develop a BCD-ADMM algorithm to jointly optimize the digital and MiLAC beamformers. The outer BCD loops alternately update $\{\mathbf u_n\}$, $\{\bm\nu_n\}$, $\{\mathbf W_n\}$, and $\{\mathbf P_n\}$. The MiLAC beamformer $\{\mathbf P_n\}$ is handled by the ADMM algorithm. Individual updates are derived below.

\subsection{Auxiliary Variables Update}
By setting the partial derivatives of the objective function \eqref{eq:max_sum_rate_wmmse_a} with respect to $u_{k,n}$ and $\nu_{k,n}$ to zero, respectively, while fixing the other variables, we obtain the updates of $\mathbf{u}_n$ and $\bm\nu_n$:
\begin{equation}
    u_{k,n}=\frac{\mathbf{h}_{k,n}^H\mathbf{P}_n\mathbf{w}_{k,n}}{\sum_{j=1}^K|\mathbf{h}_{k,n}^H\mathbf{P}_n\mathbf{w}_{j,n}|^2+\sigma^2},\label{eq:u-subproblm_solution}
\end{equation}
and
\begin{equation}
    \nu_{k,n}=(1-u_{k,n}^*\mathbf{h}_{k,n}^H\mathbf{P}_n\mathbf{w}_{k,n})^{-1},~\forall n\in\mathcal{N},~\forall k\in\mathcal{K}.\label{eq:omega-subproblm_solution}
\end{equation}

\subsection{Digital Precoder Optimization}
For fixed $\{\mathbf{u}_n\}$, $\{\bm\nu_n\}$ and $\{\mathbf{P}_n\}$, digital precoders $\{\mathbf{W}_n\}$ are updated by solving the following subproblem:
\begin{subequations}\label{eq:max_sum_rate_wmmse_W}
    \begin{align}                    \min_{\{\mathbf{W}_n\}}~& \sum_{n=1}^N\sum_{k=1}^K  \left(\mathbf{w}_{k,n}^H\mathbf{Q}_n\mathbf{w}_{k,n}-2\Re\{\mathbf{a}_{k,n}^H\mathbf{w}_{k,n}\}\right)\label{eq:max_sum_rate_wmmse_W_a}\\\mathrm{s.t.}~~&\text{\eqref{eq:max_sum_rate_g}},\label{eq:max_sum_rate_wmmse_W_b}
    \end{align}
\end{subequations}
where $\mathbf{Q}_n\triangleq\mathbf{P}_n^H\left( \sum_{j=1}^K \nu_{j,n} |u_{j,n}|^2 \mathbf{h}_{j,n} \mathbf{h}_{j,n}^H\right)\mathbf{P}_n$ and $\mathbf{a}_{k,n}\triangleq\nu_{k,n}u_{k,n}\mathbf{P}_n^H\mathbf{h}_{k,n},~\forall n\in\mathcal{N},~\forall k\in\mathcal{K}$. Problem \eqref{eq:max_sum_rate_wmmse_W} can be rewritten in matrix form as
\begin{subequations}\label{eq:max_sum_rate_wmmse_W_reformulate}
    \begin{align}                    \min_{\{\mathbf{W}_n\}}~&\sum_{n=1}^N \left(\mathrm{Tr}(\mathbf{W}_n^H\mathbf{Q}_n\mathbf{W}_n)-2\Re\{\mathrm{Tr}(\mathbf{A}_n^H\mathbf{W}_n)\}\right)\label{eq:max_sum_rate_wmmse_W_reformulate_a}\\\mathrm{s.t.}~~&\sum_{n=1}^N \mathrm{Tr}(\mathbf{W}_n^H\mathbf{R}_n\mathbf{W}_n )\leq P_\mathrm{T},\label{eq:max_sum_rate_wmmse_W_reformulate_b}
    \end{align}
\end{subequations}
where $\mathbf{A}_n=[\mathbf{a}_{1,n},\mathbf{a}_{2,n},\ldots,\mathbf{a}_{K,n}]\in\C^{N_\mathrm{RF}\times K}$, and $\mathbf{R}_n\triangleq\mathbf{P}_n^H\mathbf{P}_n,~\forall n\in\mathcal{N}$. Problem \eqref{eq:max_sum_rate_wmmse_W_reformulate} is a convex quadratically constrained quadratic program (QCQP), which can be optimally solved using the Karush-Kuhn-Tucker (KKT) conditions. 

We first consider the case in which each $\mathbf P_n$ has full column rank, so that $\mathbf R_n=\mathbf P_n^H\mathbf P_n\succ\mathbf0$. Introducing the Lagrange multiplier $\mu\geq0$ associated with \eqref{eq:max_sum_rate_wmmse_W_reformulate_b}, the KKT stationary condition yields 
\begin{equation}
    (\mathbf{Q}_n + \mu \mathbf{R}_n)\mathbf{W}_n =\mathbf{A}_n,~\forall n\in\mathcal{N},\label{eq:stationary_KKT}
\end{equation}
and the complementary slackness gives 
\begin{equation}
    \mu\left(\sum_{n=1}^{N}
    \mathrm{Tr}(\mathbf W_n^H\mathbf R_n\mathbf W_n)
    -P_{\mathrm T}\right)=0.
    \label{eq:digital_complementary_slackness}
\end{equation}
To efficiently evaluate $\{\mathbf W_n\}$ when searching for $\mu$,
we use the generalized eigendecomposition of
$(\mathbf Q_n,\mathbf R_n)$~\cite{matrix_computations}:
\begin{equation}
\mathbf Q_n\mathbf V_n
=
\mathbf R_n\mathbf V_n\bm\Sigma_n,~
\mathbf V_n^H\mathbf R_n\mathbf V_n
=
\mathbf I_{N_{\mathrm{RF}}},
\label{eq:generalized_eigenvectors}
\end{equation}
where $\bm\Sigma_n=\mathrm{diag}(\lambda_{n,1},\lambda_{n,2},\ldots,\lambda_{n,N_{\mathrm{RF}}})$ contains the generalized eigenvalues and $\mathbf V_n=[\mathbf v_{n,1},\mathbf v_{n,2}\ldots,\mathbf v_{n,N_{\mathrm{RF}}}]\in\C^{N_\mathrm{RF}\times N_\mathrm{RF}}$ contains the corresponding eigenvectors. Accordingly, $\mathbf V_n^H\mathbf Q_n\mathbf V_n=\bm\Sigma_n$. Defining $\widehat{\mathbf W}_n=\mathbf V_n^{-1}\mathbf W_n$, the KKT condition in~\eqref{eq:stationary_KKT} becomes
\begin{equation}
(\bm\Sigma_n+\mu\mathbf I_{N_{\mathrm{RF}}})
\widehat{\mathbf W}_n
=
\mathbf V_n^H\mathbf A_n.
\label{eq:digital_transformed_stationarity}
\end{equation}
Meanwhile, $\mathrm{Tr}(\mathbf W_n^H\mathbf R_n\mathbf W_n)=\|\widehat{\mathbf W}_n\|_F^2$. Let $r_{Q,n}\triangleq\operatorname{rank}(\mathbf Q_n) \leq\min\{K,N_{\mathrm{RF}}\}$ and order the eigenvalues such that $\lambda_{n,i}>0$ for $i\leq r_{Q,n}$ and $\lambda_{n,i}=0$ otherwise. For $i\leq r_{Q,n}$, $[\widehat{\mathbf W}_n(\mu)]_{i,:}=\frac{\mathbf v_{n,i}^H\mathbf A_n}{\lambda_{n,i}+\mu},~\forall n\in\mathcal{N}$. For $i>r_{Q,n}$, $\mathbf Q_n\mathbf v_{n,i}=\mathbf0$. Since $\mathbf Q_n=\sum_{k=1}^{K}\nu_{k,n}^{-1}\mathbf a_{k,n}\mathbf a_{k,n}^H$, this implies $\mathbf v_{n,i}^H\mathbf A_n=\mathbf0$. Hence, $[\widehat{\mathbf W}_n(\mu)]_{i,:}=\mathbf0$ for $\mu>0$; for $\mu=0$, we choose these rows as zero to obtain the minimum-power solution. Therefore, the total radiated power is
\begin{equation}
    P_{\mathrm W}(\mu)
    =\sum_{n=1}^{N}\|\widehat{\mathbf W}_n\|_F^2
    =\sum_{n=1}^{N}\sum_{i=1}^{r_{Q,n}}
    \frac{[\bm\Psi_n]_{i,i}}{(\lambda_{n,i}+\mu)^2}.
    \label{eq:generalized_bisection}
\end{equation}
where $\bm\Psi_n\triangleq
\mathbf V_n^H\mathbf A_n\mathbf A_n^H\mathbf V_n$. Note that $P_{\mathrm W}(\mu)$ is strictly decreasing for $\mu\geq 0$. Therefore, we set $\mu=0$ if $P_{\mathrm W}(0)\leq P_T$. Otherwise, $\mu>0$ is the unique solution of $P_{\mathrm W}(\mu)=P_{\mathrm T}$ and is obtained by bisection search. Finally, $\mathbf W_n=\mathbf V_n\widehat{\mathbf W}_n(\mu)$.

\begin{remark}
If $\mathbf P_n$ is rank-deficient, let $\mathbf P_n=\mathbf U_{P,n}\mathbf D_{P,n}\mathbf V_{P,n}^H$ be its compact singular value decomposition (SVD), with $r_{P,n}\triangleq\operatorname{rank}(\mathbf P_n)$. Since $\mathbf P_n\mathbf W_n=\mathbf P_n\mathbf V_{P,n}\mathbf V_{P,n}^H\mathbf W_n$, we can restrict without loss of optimality to $\mathbf W_n=\mathbf V_{P,n}\mathbf W_{n,r}$, where
$\mathbf W_{n,r}\in\mathbb C^{r_{P,n}\times K}$. Substituting $\mathbf W_n=\mathbf V_{P,n}\mathbf W_{n,r}$ into problem \eqref{eq:max_sum_rate_wmmse_W_reformulate} gives the matrices $ \mathbf Q_{n,r}=\mathbf V_{P,n}^H\mathbf Q_n\mathbf V_{P,n},~\mathbf R_{n,r}=\mathbf V_{P,n}^H\mathbf R_n\mathbf V_{P,n} =\mathbf D_{P,n}^{2}\succ\mathbf0$, and $\mathbf A_{n,r}=\mathbf V_{P,n}^H\mathbf A_n$. Hence, the above generalized-eigendecomposition and multiplier search apply directly in the reduced-dimensional subspace, after which $\mathbf W_n=\mathbf V_{P,n}\mathbf W_{n,r}$ is recovered.
\end{remark}

\subsection{ADMM-Based MiLAC Beamformer Optimization}
For fixed $\{\mathbf{u}_n\}$, $\{\bm\nu_n\}$, and $\{\mathbf{W}_n\}$, the MiLAC beamformers $\{\mathbf{P}_n\}$ are updated by solving the following subproblem:
\begin{subequations}\label{eq:max_sum_rate_wmmse_P}
    \begin{align}                    \min_{\{\mathbf{P}_n\},\{\mathbf{B}_n\}}~&\sum_{n=1}^N \left( \mathrm{Tr}(\mathbf{P}^H_n \mathbf{M}_n \mathbf{P}_n \mathbf{C}_n) - 2\Re\{ \mathrm{Tr}(\mathbf{D}_n\mathbf{P}_n) \} \right)\label{eq:max_sum_rate_wmmse_P_a}\\\mathrm{s.t.}~~~~~&\text{\eqref{eq:max_sum_rate_b}} - \text{\eqref{eq:max_sum_rate_f}},\label{eq:max_sum_rate_wmmse_P_b}\\
    &\sum_{n=1}^N \mathrm{Tr}(\mathbf{P}_n^H\mathbf{P}_n\mathbf{C}_n)\leq P_\mathrm{T},\label{eq:max_sum_rate_wmmse_P_c}
    \end{align}
\end{subequations}
where $\mathbf{M}_n \triangleq \sum_{k=1}^K \nu_{k,n} |u_{k,n}|^2 \mathbf{h}_{k,n} \mathbf{h}_{k,n}^H$, $\mathbf{C}_n \triangleq \mathbf{W}_n \mathbf{W}_n^H$ and $\mathbf{D}_n \triangleq \sum_{k=1}^K \nu_{k,n} u_{k,n}^* \mathbf{w}_{k,n}\mathbf{h}_{k,n}^H$. Let \begin{equation}
    \mathbf{E}_1=\left[\begin{matrix} \mathbf{I}_{N_\mathrm{RF}}\\\mathbf{0}_{N_\mathrm{T}\times N_\mathrm{RF}}\end{matrix}\right], ~\mathbf{E}_2=\left[\begin{matrix}\mathbf{0}_{N_\mathrm{T}\times N_\mathrm{RF}} ~~\mathbf{I}_{N_\mathrm{T}}\end{matrix}\right],    
\end{equation}
constraint \eqref{eq:max_sum_rate_b} can be rewritten as
\begin{equation}
    \mathbf{E}_2(\mathbf{I}_{N_\mathrm{tot}}+\jmath Z_0\mathbf{B}_n)^{-1}\mathbf{E}_1=\mathbf{P}_n,~\forall n\in\mathcal{N}.\label{eq:E_2_inverse_E1}
\end{equation}
To handle the nonlinear matrix inverse in \eqref{eq:E_2_inverse_E1}, define $\tilde{\mathbf{P}}_n\triangleq(\mathbf{I}_{N_\mathrm{tot}}+\jmath Z_0\mathbf{B}_n)^{-1}\mathbf{E}_1\in\C^{N_\mathrm{tot}\times N_\mathrm{RF}}$, then $\mathbf{P}_n=\mathbf{E}_2\tilde{\mathbf{P}}_n,~\forall n\in\mathcal{N}$, and \eqref{eq:E_2_inverse_E1} can be reformulated as\begin{equation}
    \jmath Z_0\mathbf{B}_n\tilde{\mathbf{P}}_n=\mathbf{E}_1-\tilde{\mathbf{P}}_n,\label{eq:B_n_tilde_P_n}
\end{equation} which is bilinear in $\mathbf{B}_n$ and $\tilde{\mathbf{P}}_n,~\forall n\in\mathcal{N}$. Define $\mathbf{J}\triangleq\mathbf{E}_2^H\mathbf{E}_2$, then $\mathrm{Tr}(\mathbf{P}_n^H\mathbf{P}_n\mathbf{C}_n)=\mathrm{Tr}(\tilde{\mathbf{P}}_n^H\mathbf{J}\tilde{\mathbf{P}}_n\mathbf{C}_n)$. Substituting the above transformations into \eqref{eq:max_sum_rate_wmmse_P}, problem \eqref{eq:max_sum_rate_wmmse_P} is reformulated as\begin{subequations}\label{eq:max_sum_rate_wmmse_P_reformulate_1}
    \begin{align}                    \min_{\{\tilde{\mathbf{P}}_n\},\{\mathbf{B}_n\}}~&\sum_{n=1}^N ( \mathrm{Tr}(\tilde{\mathbf{P}}^H_n\tilde{\mathbf{M}}_n\tilde{\mathbf{P}}_n \mathbf{C}_n) - 2\Re\{ \mathrm{Tr}(\tilde{\mathbf{D}}_n\tilde{\mathbf{P}}_n) \} )\label{eq:max_sum_rate_wmmse_P_reformulate_1_a}\\ \mathrm{s.t.}~~~~~&\jmath Z_0\mathbf{B}_n\tilde{\mathbf{P}}_n=\mathbf{E}_1-\tilde{\mathbf{P}}_n,\label{eq:max_sum_rate_wmmse_P_reformulate_1_b}\\
    &\text{\eqref{eq:max_sum_rate_c}} - \text{\eqref{eq:max_sum_rate_f}},\label{eq:max_sum_rate_wmmse_P_reformulate_1_c}\\
    &\sum_{n=1}^N \mathrm{Tr}(\tilde{\mathbf{P}}_n^H\mathbf{J}\tilde{\mathbf{P}}_n\mathbf{C}_n)\leq P_\mathrm{T},\label{eq:max_sum_rate_wmmse_P_reformulate_1_d}
    \end{align}
\end{subequations}
where $\tilde{\mathbf{M}}_n=\mathbf{E}_2^H \mathbf{M}_n \mathbf{E}_2$ and $\tilde{\mathbf{D}}_n=\mathbf{D}_n\mathbf{E}_2,~\forall n\in\mathcal{N}$. To solve problem \eqref{eq:max_sum_rate_wmmse_P_reformulate_1}, we apply an ADMM framework to handle the bilinear constraint \eqref{eq:max_sum_rate_wmmse_P_reformulate_1_b}. Let $\mathbf{U}_n\in\mathbb{C}^{N_\mathrm{tot}\times N_\mathrm{RF}}$ denote the scaled Lagrange
multiplier associated with \eqref{eq:max_sum_rate_wmmse_P_reformulate_1_b} at subcarrier $n$, and let $\rho>0$ denote the penalty parameter. The augmented Lagrangian function in the scaled form is given by \cite{boyd_admm}
\begin{equation}
\begin{aligned}
&\mathcal{L}_{\rho}
\left(
\{\tilde{\mathbf{P}}_n\},
\{\mathbf{B}_n\},
\{\mathbf{U}_n\}
\right)\\
&=
\sum_{n=1}^{N}
(
\mathrm{Tr}(
\tilde{\mathbf{P}}_n^H
\tilde{\mathbf{M}}_n
\tilde{\mathbf{P}}_n
\mathbf{C}_n)
-
2\Re\{
\mathrm{Tr}(
\tilde{\mathbf{D}}_n
\tilde{\mathbf{P}}_n)\}
)
\\
&+
\rho
\sum_{n=1}^{N}
\left\|
\jmath Z_0\mathbf{B}_n\tilde{\mathbf{P}}_n-\mathbf{E}_1+ \tilde{\mathbf{P}}_n
+
\mathbf{U}_n
\right\|_F^2
-
\rho
\sum_{n=1}^{N}
\left\|
\mathbf{U}_n
\right\|_F^2,
\end{aligned}
\label{eq:augmented_lagrangian_P}
\end{equation}

At each ADMM iteration, we update $\{\tilde{\mathbf{P}}_n\}$, $\{\mathbf{B}_n\}$, and the dual variables $\{\mathbf{U}_n\}$ alternately until convergence. The detailed algorithm is given below.

\subsubsection{$\{\tilde{\mathbf P}_n\}$ Update} 
In each ADMM iteration, $\{\tilde{\mathbf{P}}_n\}$ is updated by solving the following subproblem 
\begin{subequations}\label{eq:update_tilde_Pn}
    \begin{align}
    \min_{\{\tilde{\mathbf{P}}_n\}}~&\sum_{n=1}^N \left( \mathrm{Tr}(\tilde{\mathbf{P}}^H_n\tilde{\mathbf{M}}_n\tilde{\mathbf{P}}_n \mathbf{C}_n) - 2\Re\{ \mathrm{Tr}(\tilde{\mathbf{D}}_n\tilde{\mathbf{P}}_n) \} \right)\\
    &+
    \rho
    \sum_{n=1}^{N}
    \left\|
    \tilde{\mathbf{B}}_n\tilde{\mathbf{P}}_n-\mathbf{E}_1
    +
    \mathbf{U}_n
    \right\|_F^2,\label{eq:update_tilde_Pn_a}\\
    \mathrm{s.t.}~~
    &\sum_{n=1}^N \mathrm{Tr}(\tilde{\mathbf{P}}_n^H\mathbf{J}\tilde{\mathbf{P}}_n\mathbf{C}_n)\leq P_\mathrm{T},\label{eq:update_tilde_Pn_b}
    \end{align}
\end{subequations}
where $\tilde{\mathbf{B}}_n\triangleq\mathbf{I}_{N_\mathrm{tot}}+\jmath Z_0\mathbf{B}_n$. Subproblem \eqref{eq:update_tilde_Pn} is a convex QCQP. Compared to \eqref{eq:max_sum_rate_wmmse_W_reformulate}, this subproblem has a more involved matrix structure because the objective and constraint contain the two-sided quadratic term involving $\tilde{\mathbf M}_n$, $\mathbf{J}$, and $\mathbf C_n$. Nevertheless, the optimal solution can still be characterized through the KKT conditions. Introducing the Lagrange multiplier $\kappa\geq 0$ associated with \eqref{eq:update_tilde_Pn_b}, the KKT stationary condition yields
\begin{equation}
    (\tilde{\mathbf{M}}_n+\kappa\mathbf{J})
\tilde{\mathbf{P}}_n
\mathbf{C}_n
+
\rho
\tilde{\mathbf{B}}_n^H
\tilde{\mathbf{B}}_n
\tilde{\mathbf{P}}_n
=
\tilde{\mathbf{D}}_n^H
+
\rho
\tilde{\mathbf{B}}_n^H
\left(
\mathbf{E}_1
-
\mathbf{U}_n
\right).\label{eq:stationary_tilde_P}
\end{equation}
Equation \eqref{eq:stationary_tilde_P} is a generalized Sylvester equation. A direct approach is to vectorize it and solve a linear system of dimension $N_{\mathrm{tot}}N_{\mathrm{RF}}$ for each subcarrier and each candidate value of $\kappa$. Although this approach allows $\kappa$ to be determined by bisection, repeatedly solving these systems can incur substantial computational cost. We thus develop a reduced-complexity approach that, after preprocessing, determines $\kappa$ through a bisection search requiring only scalar power evaluations.

We first apply the eigenvalue decomposition (EVD) of $\mathbf C_n$ as $\mathbf{C}_n=\mathbf{T}_n\mathbf{\Xi}_n\mathbf{T}_n^H$, where $\mathbf{T}_n\in\C^{N_{\mathrm{RF}}\times N_{\mathrm{RF}}}$ is unitary and $\mathbf{\Xi}_n=\mathrm{diag}(\xi_{n,1},\xi_{n,2},\ldots,\xi_{n,N_{\mathrm{RF}}})$. Define $\tilde{\tilde{\mathbf{P}}}_n\triangleq
\tilde{\mathbf{P}}_n\mathbf{T}_n$, and $\tilde{\tilde{\mathbf{B}}}_n
\triangleq
(
\tilde{\mathbf{D}}_n^H
+
\rho
\tilde{\mathbf{B}}_n^H(\mathbf{E}_1-\mathbf{U}_n))
\mathbf{T}_n$. Right-multiplying \eqref{eq:stationary_tilde_P} by $\mathbf{T}_n$ gives
\begin{equation}
(\tilde{\mathbf M}_n+\kappa\mathbf J)
\tilde{\tilde{\mathbf{P}}}_n\mathbf\Xi_n
+
\rho\tilde{\mathbf{B}}_n^H\tilde{\mathbf{B}}_n\tilde{\tilde{\mathbf{P}}}_n
=\tilde{\tilde{\mathbf{B}}}_n,~\forall n\in\mathcal N.
\label{eq:stationary_tilde_tilde_P}
\end{equation}
Since $\mathbf{\Xi}_n$ is diagonal, \eqref{eq:stationary_tilde_tilde_P} can be decomposed into $N_{\mathrm{RF}}$ linear systems as\begin{equation}
\left(\xi_{n,i}(\tilde{\mathbf{M}}_n+\kappa\mathbf{J})+\rho\tilde{\mathbf{B}}_n^H\tilde{\mathbf{B}}_n
\right)\tilde{\tilde{\mathbf{p}}}_{n,i}
=\tilde{\tilde{\mathbf{b}}}_{n,i},
\label{eq:tilde_tildeP_linear_system}
\end{equation}
for $i=1,2,\ldots,N_{\mathrm{RF}}$, where $\tilde{\tilde{\mathbf{p}}}_{n,i}\in\C^{N_\mathrm{tot}\times 1}$ and $\tilde{\tilde{\mathbf{b}}}_{n,i}\in\C^{N_\mathrm{tot}\times 1}$ denote the $i$-th column of $\tilde{\tilde{\mathbf{P}}}_n$ and $\tilde{\tilde{\mathbf{B}}}_n$, respectively. To avoid repeatedly solving these systems for different values of $\kappa$, define $\mathbf{G}_{n,i}
\triangleq
\xi_{n,i}\tilde{\mathbf{M}}_n+\rho\tilde{\mathbf{B}}_n^H\tilde{\mathbf{B}}_n$ and $\tilde{\mathbf{J}}_{n,i}\triangleq\xi_{n,i}\mathbf{J}$. Then, \eqref{eq:tilde_tildeP_linear_system} can be rewritten as
\begin{equation}
(\mathbf{G}_{n,i}+\kappa\tilde{\mathbf{J}}_{n,i})
\tilde{\tilde{\mathbf{p}}}_{n,i}
=\tilde{\tilde{\mathbf{b}}}_{n,i},~\forall n\in\mathcal{N}.
\label{eq:tilde_tildeP_linear_system_compact}
\end{equation}
Since $\tilde{\mathbf{B}}_n$ is nonsingular and $\rho>0$, $\mathbf{G}_{n,i}\succ\mathbf{0}$ and $\tilde{\mathbf{J}}_{n,i}\succeq\mathbf{0}$. We therefore use the generalized eigendecomposition of
$(\tilde{\mathbf J}_{n,i},\mathbf G_{n,i})$
\cite{matrix_computations}:
\begin{equation}
\tilde{\mathbf J}_{n,i}\mathbf V_{n,i}^{\mathrm{P}}
=
\mathbf G_{n,i}\mathbf V_{n,i}^{\mathrm{P}}\bm\Delta_{n,i},
~
(\mathbf V_{n,i}^{\mathrm{P}})^H
\mathbf G_{n,i}\mathbf V_{n,i}^{\mathrm{P}}
=
\mathbf I_{N_{\mathrm{tot}}},
\label{eq:milac_generalized_eigenvalue}
\end{equation}
where
$\bm\Delta_{n,i}
=\mathrm{diag}(d_{n,i,1},d_{n,i,2},\ldots,d_{n,i,N_{\mathrm{tot}}})$ contains generalized eigenvalues $d_{n,i,j}\geq0$, and $\mathbf{V}_{n,i}^{\mathrm{P}}\in\C^{N_{\mathrm{tot}}\times N_{\mathrm{tot}}}$ contains the corresponding eigenvectors, for $i=1,2,\ldots,N_{\mathrm{RF}}$. Using \eqref{eq:milac_generalized_eigenvalue}, we obtain
\begin{align}
\mathbf{G}_{n,i}
+\kappa\tilde{\mathbf{J}}_{n,i}
&=
((\mathbf{V}_{n,i}^{\mathrm{P}})^H)^{-1}
(\mathbf{I}_{N_{\mathrm{tot}}}
+\kappa\bm\Delta_{n,i})
(\mathbf{V}_{n,i}^{\mathrm{P}})^{-1},\label{eq:tilde_tildeP}
\end{align}
for $i=1,2,\ldots,N_{\mathrm{RF}}$. Substituting \eqref{eq:tilde_tildeP} into \eqref{eq:tilde_tildeP_linear_system_compact} gives
\begin{equation}
\tilde{\tilde{\mathbf{p}}}_{n,i}(\kappa)
=
\mathbf{V}_{n,i}^{\mathrm{P}}
\left(
\mathbf{I}_{N_{\mathrm{tot}}}
+\kappa\bm\Delta_{n,i}
\right)^{-1}
(\mathbf{V}_{n,i}^{\mathrm{P}})^H
\tilde{\tilde{\mathbf{b}}}_{n,i}.
\label{eq:solution_tilde_tildeP}
\end{equation}
Using $\mathbf{C}_n=\mathbf{T}_n\mathbf{\Xi}_n\mathbf{T}_n^H$ and $\tilde{\tilde{\mathbf{P}}}_n=\tilde{\mathbf{P}}_n\mathbf{T}_n$, the total radiated power can be expressed as
\begin{align}
P_{\mathrm{rad}}(\kappa)=
\sum_{n=1}^{N}
\mathrm{Tr}
(
\tilde{\tilde{\mathbf{P}}}_n^H
\mathbf{J}
\tilde{\tilde{\mathbf{P}}}_n
\mathbf{\Xi}_n
)
=
\sum_{n=1}^{N}
\sum_{i=1}^{N_{\mathrm{RF}}}
(\tilde{\tilde{\mathbf{p}}}_{n,i})^H
\tilde{\mathbf{J}}_{n,i}
\tilde{\tilde{\mathbf{p}}}_{n,i}.
\label{eq:radiated_power_tildeP_1}
\end{align}
Define $\mathbf{a}_{n,i}^{\mathrm{P}}
\triangleq(\mathbf{V}_{n,i}^{\mathrm{P}})^H\tilde{\tilde{\mathbf{b}}}_{n,i}$, substituting \eqref{eq:solution_tilde_tildeP} into
\eqref{eq:radiated_power_tildeP_1} yields
\begin{align}
&P_{\mathrm{rad}}(\kappa)\notag\\
&=
\sum_{n=1}^{N}
\sum_{i=1}^{N_{\mathrm{RF}}}
(\mathbf{a}_{n,i}^{\mathrm{P}})^H
(\mathbf{I}+\kappa\bm\Delta_{n,i})^{-1}
\bm\Delta_{n,i}(
\mathbf{I}+\kappa\bm\Delta_{n,i})^{-1}
\mathbf{a}_{n,i}^{\mathrm{P}}
\notag\\
&=
\sum_{n=1}^{N}
\sum_{i=1}^{N_{\mathrm{RF}}}
\sum_{j=1}^{N_{\mathrm{tot}}}
\frac{
d_{n,i,j}
\left|[\mathbf{a}_{n,i}^{\mathrm{P}}]_j\right|^2
}{
\left(1+\kappa d_{n,i,j}\right)^2
}.
\label{eq:radiated_power_kappa}
\end{align}
$P_{\mathrm{rad}}(\kappa)$ is strictly decreasing for $\kappa\geq0$. Hence, if $P_{\mathrm{rad}}(0)\leq P_{\mathrm T}$, we set $\kappa=0$; otherwise, the unique positive solution of $P_{\mathrm{rad}}(\kappa)=P_{\mathrm T}$ is determined by bisection. Then we compute ${\tilde{\tilde{\mathbf p}}_{n,i}}$ using \eqref{eq:solution_tilde_tildeP} for all $i$ and recover $\tilde{\mathbf P}_n=\tilde{\tilde{\mathbf P}}_n\mathbf T_n^H,~\forall n\in\mathcal{N}$.

\begin{remark}
A direct evaluation of \eqref{eq:stationary_tilde_P} for each candidate $\kappa$ can be obtained by vectorization, which requires solving a linear system of dimension $N_{\mathrm{tot}}N_{\mathrm{RF}}$ for each subcarrier. This results in a complexity of $\mathcal{O}\left(I_\kappa N(N_{\mathrm{tot}}N_{\mathrm{RF}})^3\right)$ over $I_\kappa$ bisection iterations. With the proposed computation, the EVDs of $\{\mathbf C_n\}$ and the generalized eigendecompositions in \eqref{eq:milac_generalized_eigenvalue} are performed only once per $\{\tilde{\mathbf P}_n\}$ update, with complexity $\mathcal{O}\left(NN_{\mathrm{RF}}^3+NN_{\mathrm{RF}}N_{\mathrm{tot}}^3\right)$. After this preprocessing, the complexity of bisection in \eqref{eq:radiated_power_kappa} is only $\mathcal O(I_\kappa NN_{\mathrm{RF}}N_{\mathrm{tot}})$. Hence, the repeated $N_{\mathrm{tot}}N_{\mathrm{RF}}$-dimensional matrix factorizations during the search for $\kappa$ are replaced by scalar evaluations.
\end{remark}

\subsubsection{$\{\mathbf{B}_n\}$ Update} In each ADMM iteration, $\{\mathbf{B}_n\}$ is updated by solving the following subproblem  
\begin{subequations}\label{eq:update_Bn_1}
    \begin{align} \min_{\{\mathbf{B}_n\}}~&\sum_{n=1}^{N}
\left\|
\jmath Z_0\mathbf{B}_n\tilde{\mathbf{P}}_n
-(
\mathbf{E}_1-\tilde{\mathbf{P}}_n-\mathbf{U}_n
)\right\|_F^2\label{eq:update_Bn_1_a}\\\mathrm{s.t.}~~&\text{\eqref{eq:max_sum_rate_c}} - \text{\eqref{eq:max_sum_rate_f}},\label{eq:update_Bn_1_b}
    \end{align}
\end{subequations}
Define $\breve{\mathbf{P}}_n\triangleq\left[
\Re\{\jmath Z_0\tilde{\mathbf{P}}_n\},
\Im\{\jmath Z_0\tilde{\mathbf{P}}_n\}
\right]
\in\mathbb{R}^{N_\mathrm{tot}\times 2N_\mathrm{RF}}$, and $\bm\Gamma_{n}\triangleq
\left[
\Re\{\mathbf{E}_1-\tilde{\mathbf{P}}_n-\mathbf{U}_n\},
\Im\{\mathbf{E}_1-\tilde{\mathbf{P}}_n-\mathbf{U}_n\}
\right]\in\mathbb{R}^{N_\mathrm{tot}\times 2N_\mathrm{RF}}$. Then \eqref{eq:update_Bn_1} is equivalent to
\begin{subequations}\label{eq:update_Bn_2}
    \begin{align} \min_{\{\mathbf{B}_n\}}
~&\sum_{n=1}^{N}
\left\|
\mathbf{B}_n\breve{\mathbf{P}}_n
-
\bm\Gamma_n
\right\|_F^2\label{eq:update_Bn_2_a}\\\mathrm{s.t.}~~&\text{\eqref{eq:max_sum_rate_c}} - \text{\eqref{eq:max_sum_rate_f}}.\label{eq:update_Bn_2_b}
    \end{align}
\end{subequations}
To incorporate the symmetric constraint \eqref{eq:max_sum_rate_c} and the mapping constraint \eqref{eq:max_sum_rate_d}, let $\bar{\mathbf b}_n\in\mathbb R^{\tilde N_{\mathrm{tot}}\times1}$ collect all tunable susceptance components $B_{m,m',n}$, $1\leq m\leq m'\leq N_{\mathrm{tot}}$, in the same order as $\mathbf b_{\mathrm c}$. The linear model in \eqref{eq:B_linear_model} gives
\begin{equation}
\bar{\mathbf b}_n
=
F_1(\omega_n)\mathbf b_{\mathrm c}
+F_2(\omega_n)\mathbf1_{\tilde N_{\mathrm{tot}}},~\forall n\in\mathcal N.
\label{eq:b_n_linear_model}
\end{equation}
Let $\breve{\mathbf P}_n^T
=[\bar{\mathbf a}_{n,1},\bar{\mathbf{a}}_{n,2},\ldots,\bar{\mathbf a}_{n,N_{\mathrm{tot}}}]$, where $\bar{\mathbf a}_{n,i}\in\mathbb R^{2N_{\mathrm{RF}}\times1}$, and define $\bm\gamma_n\triangleq\mathrm{vec}(\bm\Gamma_n^T),~\forall n\in\mathcal{N}$. We construct a real matrix $\bar{\mathbf{A}}_n\in\mathbb{R}^{2N_\mathrm{tot}N_\mathrm{RF}\times \tilde{N}_\mathrm{tot}}$, with $\bar{\mathbf{A}}_{n,i,j}\in\mathbb{R}^{2N_\mathrm{RF}\times (N_\mathrm{tot}-j+1)}$ indicating the $i$-th block row and $j$-th block column. Each block $\bar{\mathbf{A}}_{n,i,j}$ is defined as
\begin{equation}
[\bar{\mathbf{A}}_{n,i,j}]_{:,q}=
\begin{cases}
\bar{\mathbf{a}}_{n,j},
&\text{if}~~ i=j, q=1;\\
\bar{\mathbf{a}}_{n,j}-\bar{\mathbf{a}}_{n,j+q-1},
&\text{if}~~ i=j, q\geq 2;\\
\bar{\mathbf{a}}_{n,j+q-1}-\bar{\mathbf{a}}_{n,j},
&\text{if}~~ i=j+q-1, q\geq 2;\\
\mathbf{0}_{2N_\mathrm{RF}\times 1},
&\text{otherwise},
\end{cases}
\label{eq:Abar_block_definition_proof}
\end{equation}
for $i,j=1,2,\ldots,N_\mathrm{tot}$ and $q=1,2,\ldots,N_\mathrm{tot}-j+1$. This construction accounts for the diagonal and off-diagonal contributions of each component in \eqref{eq:max_sum_rate_d}, giving $\bar{\mathbf A}_n\bar{\mathbf b}_n=\mathrm{vec}\left((\mathbf B_n\breve{\mathbf P}_n)^T\right),~\forall n\in\mathcal{N}$. Consequently,
\begin{equation}
\|\mathbf B_n\breve{\mathbf P}_n-\bm\Gamma_n\|_F^2
=
\|\bar{\mathbf A}_n\bar{\mathbf b}_n
-\bm\gamma_n\|_2^2.
\label{eq:BM_to_Ax_proof_identity}
\end{equation}
Thus, the symmetry constraint \eqref{eq:max_sum_rate_c} and component-mapping constraint \eqref{eq:max_sum_rate_d} are incorporated into $\bar{\mathbf A}_n$. Substituting \eqref{eq:b_n_linear_model} into
\eqref{eq:BM_to_Ax_proof_identity}, define $\tilde{\mathbf{A}}
    \triangleq
    \left[
    F_1(\omega_1)\bar{\mathbf{A}}_1^T,
    \ldots,
    F_1(\omega_N)\bar{\mathbf{A}}_N^T
    \right]^T,
    ~\tilde{\bm\gamma}
    \triangleq\big[
    (\bm\gamma_1
    -F_2(\omega_1)\bar{\mathbf{A}}_1\mathbf{1}_{\tilde{N}_\mathrm{tot}})^T,
    \ldots,
    (\bm\gamma_N
    -F_2(\omega_N)\bar{\mathbf{A}}_N\mathbf{1}_{\tilde{N}_\mathrm{tot}})^T
\big]^T$. Problem \eqref{eq:update_Bn_1} is therefore equivalent to the following linearly constrained least-squares problem
\begin{subequations}\label{eq:update_bc_box_ls}
    \begin{align}
    \min_{\mathbf{b}_\mathrm{c}}~~&\left\|\tilde{\mathbf{A}}\mathbf{b}_\mathrm{c}
    -
    \tilde{\bm\gamma}
    \right\|_2^2
    \label{eq:update_bc_box_ls_a}\\
    \mathrm{s.t.}~~&B_\mathrm{min}\mathbf1_{\tilde N_{\mathrm{tot}}}\leq\mathbf b_{\mathrm c}\leq
B_\mathrm{max}\mathbf1_{\tilde N_{\mathrm{tot}}}.
    \label{eq:update_bc_box_ls_b}
    \end{align}
\end{subequations}
To efficiently solve \eqref{eq:update_bc_box_ls}, we adopt PGD. Let $\mathcal B_{\mathrm c}
\triangleq[B_{\min},B_{\max}]^{\tilde N_{\mathrm{tot}}}$ denote the feasible set. Then $\mathbf{b}_\mathrm{c}$ can be updated by PGD as 
\begin{equation}
    \mathbf{b}_\mathrm{c}^{(t+1)} = \mathrm{Proj}_{\mathcal{B}_\mathrm{c}} \left( \mathbf{b}_\mathrm{c}^{(t)} - \zeta \mathrm{Grad}(\mathbf{b}_\mathrm{c}^{(t)}) \right),\label{eq:bc_PGD_update}
\end{equation}
where $\mathbf{b}_\mathrm{c}^{(t)}$ denotes the value of $\mathbf{b}_\mathrm{c}$ at the $t$-th iteration, $\mathrm{Grad}(\mathbf{b}_\mathrm{c})=2\tilde{\mathbf{A}}^T(\tilde{\mathbf{A}}\mathbf{b}_\mathrm{c}-\tilde{\bm\gamma})$, $\mathrm{Proj}_{\mathcal{B}_\mathrm{c}}$ represents the projection onto set $\mathcal{B}_\mathrm{c}$, and $\zeta>0$ is the step size. For any vector $\mathbf{x}\in\mathbb{R}^{\tilde{N}_\mathrm{tot}\times 1}$, $\mathrm{Proj}_{\mathcal{B}_\mathrm{c}}(\mathbf{x})$ denotes the element-wise projection onto $\mathcal{B}_\mathrm{c}$, given by
\begin{equation}
    [\mathrm{Proj}_{\mathcal{B}_\mathrm{c}}(\mathbf{x})]_i
    =\min\left\{
    B_\mathrm{max},
    \max\left\{
    B_\mathrm{min},[\mathbf{x}]_i
    \right\}
    \right\},
\end{equation}
for $i=1,2,\ldots,\tilde{N}_\mathrm{tot}$. Since $2\tilde{\mathbf A}^T\tilde{\mathbf A}
\succeq\mathbf0$, problem \eqref{eq:update_bc_box_ls} is convex, and its gradient has Lipschitz constant $L_G=2\|\tilde{\mathbf A}\|_2^2$. A step size of $\zeta=1/L_G$ ensures that the convergence of PGD to a global minimizer of \eqref{eq:update_bc_box_ls}. After obtaining $\mathbf{b}_\mathrm{c}$, we recover $\{\bar{\mathbf{b}}_n\}$ using \eqref{eq:b_n_linear_model} and reconstruct $\{\mathbf{B}_n\}$ through \eqref{eq:max_sum_rate_d}.

\subsubsection{$\{\mathbf{U}_n\}$ Update} In each ADMM iteration, dual variables $\{\mathbf{U}_n\}$ are updated as
\begin{equation}
    \mathbf{U}_n
    =
    \mathbf{U}_n
    +
    \tilde{\mathbf{B}}_n\tilde{\mathbf{P}}_n
    -
    \mathbf{E}_1,
    ~\forall n\in\mathcal{N},
    \label{eq:U_update}
\end{equation}
where $\tilde{\mathbf{B}}_n=\mathbf{I}_{N_\mathrm{tot}}+\jmath Z_0\mathbf{B}_n$. The proposed BCD-ADMM algorithm is summarized in
Algorithm~\ref{alg:wmmse_admm_pgd}.

\begin{algorithm}[t]
    \caption{Proposed BCD-ADMM Algorithm for Wideband Hybrid Digital-MiLAC Beamforming}
    \label{alg:wmmse_admm_pgd}
    \begin{algorithmic}[1]
        \REQUIRE $\{\mathbf{h}_{k,n}\}$, $\sigma^2$, $P_\mathrm{T}$, $Z_0$, $B_\mathrm{min}$, $B_\mathrm{max}$, $\{F_1(\omega_n),F_2(\omega_n)\}$, and $\rho$.
        \ENSURE $\{\mathbf{W}_n^\star\}$ and $\{\mathbf{P}_n^\star\}$.
        \STATE Initialize $\mathbf{b}_\mathrm{c}$, construct $\{\mathbf{B}_n\}$ by \eqref{eq:b_n_linear_model} and \eqref{eq:max_sum_rate_d}, recover $\{\mathbf{P}_n\}$ using \eqref{eq:max_sum_rate_b}, and initialize $\{\mathbf{W}_n\}$ satisfying \eqref{eq:max_sum_rate_g}.

        \WHILE{the objective in \eqref{eq:max_sum_rate_wmmse_a} does not converge}
            \STATE Update $\{u_{k,n}\}$ and $\{\nu_{k,n}\}$ using \eqref{eq:u-subproblm_solution} and \eqref{eq:omega-subproblm_solution}.
            \STATE Calculate $\{\tilde{\mathbf{P}}_n\}$ from $\{\mathbf{B}_n\}$, and initialize $\{\mathbf{U}_n\}$.

            \WHILE{the objective in \eqref{eq:max_sum_rate_wmmse_P_reformulate_1_a} does not converge}
                \STATE Set $\kappa=0$ if
                $P_{\mathrm{rad}}(0)\leq P_{\mathrm T}$; otherwise determine $\kappa>0$
                by bisection using
                \eqref{eq:radiated_power_kappa}. Update $\{\tilde{\tilde{\mathbf p}}_{n,i}\}$ using \eqref{eq:solution_tilde_tildeP},
                and recover $\tilde{\mathbf P}_n
                =\tilde{\tilde{\mathbf P}}_n\mathbf T_n^H,~\forall n\in\mathcal N$.
                \STATE Compute $\{\breve{\mathbf{P}}_n,\bm\Gamma_n\}$, and construct $\tilde{\mathbf{A}}$ and $\tilde{\bm\gamma}$.
                \STATE Solve \eqref{eq:update_bc_box_ls} by PGD using the update in \eqref{eq:bc_PGD_update}.
                \STATE Recover $\{\bar{\mathbf{b}}_n\}$ from \eqref{eq:b_n_linear_model} and $\{\mathbf{B}_n\}$ by \eqref{eq:max_sum_rate_d}.
                \STATE Update $\{\tilde{\mathbf{B}}_n\}$ as $\tilde{\mathbf{B}}_n=\mathbf{I}_{N_\mathrm{tot}}+\jmath Z_0\mathbf{B}_n,~\forall n\in\mathcal{N}$.
                \STATE Update $\{\mathbf{U}_n\}$ by \eqref{eq:U_update}.
            \ENDWHILE
            \STATE Reconstruct $\{\mathbf{P}_n\}$ by \eqref{eq:max_sum_rate_b}.
            \STATE Set $\mu=0$ if
            $P_{\mathrm W}(0)\leq P_{\mathrm T}$;
            otherwise determine $\mu>0$ by bisection using \eqref{eq:generalized_bisection}. Update $\{\mathbf{W}_n\}$ using $\mathbf W_n=\mathbf V_n\widehat{\mathbf W}_n(\mu)$.
        \ENDWHILE
        \STATE Return $\{\mathbf{W}_n\}$ and $\{\mathbf{P}_n\}$ as $\{\mathbf{W}_n^\star\}$ and $\{\mathbf{P}_n^\star\}$.
    \end{algorithmic}
\end{algorithm}

\section{Simulations}
\label{sec:simulations}
In this section, we evaluate the frequency-dependent hybrid digital-MiLAC beamforming design and compare it with several benchmark schemes through numerical simulations.

\subsection{Simulation Settings}
In the considered MU-MISO-OFDM system, unless otherwise specified, we set the bandwidth $B$ to 100 MHz, the number of subcarriers, transmit antennas, and users to $N=64$, $N_\text{T}=64$, and $K=4$, respectively. The OFDM channel of user $k$ is modeled by $D$ independent time-domain taps $\{\bar{\mathbf h}_{k,d}\}_{d=0}^{D-1}$, with $D=16$, where $\bar{\mathbf h}_{k,d}\sim\mathcal{CN}(\mathbf0,p_d\mathbf I_{N_{\mathrm T}})$. We adopt the normalized PDP given by $p_d=\frac{e^{-d/\epsilon}}{\sum_{i=0}^{D-1}e^{-i/\epsilon}},~d=0,1,\ldots,D-1$. The frequency-domain channel $\mathbf{h}_{k,n}\in\C^{N_\mathrm{T}\times 1}$ is obtained by $\mathbf{h}_{k,n}=\sum_{d=0}^{D-1}\bar{\mathbf{h}}_{k,d}e^{-\jmath \frac{2\pi}{N}(n-1)d}$. To characterize frequency selectivity, we use the normalized root mean square (RMS) delay spread $\tau_{\mathrm{rms}}/T_s$, where $T_s=1/B$ is the sampling period and $\tau_{\mathrm{rms}}=T_s\sqrt{\sum_{d=0}^{D-1} p_d\left(d-\bar{d}\right)^2}$ with $\bar d=\sum_{d=0}^{D-1} p_d d$. The noise power is normalized to $\sigma^2=1$ for all users and subcarriers. The signal-to-noise ratio (SNR) is defined as $\frac{P_{\mathrm T}}{N\sigma^2}$.

\subsection{Benchmark Schemes}
We use \textbf{HDM} and \textbf{HDA} to denote hybrid digital-MiLAC and hybrid digital-analog schemes, respectively, and \textbf{FD} and \textbf{FI} to denote frequency-dependent and
frequency-independent responses, respectively. The proposed wideband design using the BCD-ADMM
algorithm is denoted by \textbf{FD-HDM}. For digital beamforming, define the subcarrier sum-rate as $R_n(\frac{1}{2}\mathbf W_n^\mathrm{D})\triangleq
\sum_{k=1}^{K}
\log_2\left(
1+\frac{|\frac{1}{2}\mathbf h_{k,n}^H\mathbf w_{k,n}^\mathrm{D}|^2}
{\sum_{j\neq k}|\frac{1}{2}\mathbf h_{k,n}^H\mathbf w_{j,n}^\mathrm{D}|^2
+\sigma^2}
\right)$,\footnote{The appearance of the $1/2$ factor is for the same reason as described in Section~\ref{subsec:milac_visualization}.} where $\mathbf W_n^\mathrm{D}=[\mathbf w_{1,n}^\mathrm{D},\mathbf w_{2,n}^\mathrm{D},\ldots,\mathbf w_{K,n}^\mathrm{D}]$. For hybrid beamforming, define $R_n(\mathbf{P}_n,\mathbf W_n)\triangleq
\sum_{k=1}^{K}
\log_2\left(
1+\frac{|\mathbf h_{k,n}^H\mathbf{P}_n\mathbf w_{k,n}|^2}
{\sum_{j\neq k}|\mathbf h_{k,n}^H\mathbf{P}_n\mathbf w_{j,n}|^2
+\sigma^2}
\right)$. We compare FD-HDM with the following schemes.

\subsubsection{Digital Beamforming} The digital precoders $\{\mathbf W_n^\mathrm{D}\}$ are designed across all subcarriers using WMMSE~\cite{WMMSE} to solve
\begin{equation}
\{(\mathbf W_n^\mathrm{D})^\sharp\}
\in
\operatorname*{\arg~\max}_{\sum_{n=1}^{N}\|\frac{1}{2}\mathbf W_n^\mathrm{D}\|_F^2
    \leq P_{\mathrm T}}
\frac{1}{N}\sum_{n=1}^{N}R_n\left(\frac{1}{2}\mathbf W_n^\mathrm{D}\right).
\label{eq:benchmark_digital}
\end{equation}

\subsubsection{FI-HDM} For the \textbf{FI-HDM} scheme, MiLAC beamformers are designed such that they remain the same across all subcarriers, i.e., $\mathbf P_n=\mathbf P,~\mathbf B_n=\mathbf B$, and $B_{m,m',n}=B_{m,m'},~n\in\mathcal N$. The joint design problem is given by
    \begin{equation}
\begin{aligned}
&(\{B_{m,m'}^\sharp\},\{\mathbf W_n^\sharp\})\\
&\in
\operatorname*{\arg~\max}_{\substack{
    \text{\eqref{eq:max_sum_rate_b}-\eqref{eq:max_sum_rate_d}},~B_{m,m'}\in[B_{\mathrm{min}},B_{\mathrm{max}}],~\forall m,m',\\
    \sum_{n=1}^{N}
    \|\mathbf P^{\mathrm{FI}}\mathbf W_n\|_F^2
    \leq P_{\mathrm T}
}}
\frac{1}{N}\sum_{n=1}^{N}
R_n\left(\mathbf P,\mathbf W_n\right).
\end{aligned}
\label{eq:benchmark_fi_hdm}
\end{equation}
We solve \eqref{eq:benchmark_fi_hdm} using the proposed BCD-ADMM algorithm with $F_1(\omega_n)=1$ and $F_2(\omega_n)=0,~\forall n\in\mathcal N$. $\{B_{m,m'}^\sharp\}$ are substituted into \eqref{eq:B_linear_model} to obtain the physical FD responses $\{\mathbf P_n^\mathrm{FI}\}$. With $\{\mathbf P_n^\mathrm{FI}\}$ fixed, digital precoders are optimized to obtain $\{\mathbf{W}_n^\mathrm{FI}\}$ \cite{WMMSE}. Finally, $\{\mathbf P_n^\mathrm{FI}\}$ and $\{\mathbf{W}_n^\mathrm{FI}\}$ are plugged into \eqref{eq:max_sum_rate_a} to evaluate average sum-rate.

\subsubsection{Proj-HDM} For the \textbf{Proj-HDM} scheme, a FI MiLAC beamformer $\mathbf{P}$ is first optimized using the method in \cite{wideband_fi_milac_yiyang}. It is then projected onto physical MiLAC constraints by solving
\begin{equation}
\{(\mathbf P_n^\mathrm{Proj})^\sharp\}
\in
\operatorname*{\arg~\min}_{
    \text{\eqref{eq:max_sum_rate_b}-\eqref{eq:max_sum_rate_f}}}
\sum_{n=1}^{N}
\left\|
\mathbf P_n^\mathrm{Proj}
-\mathbf P
\right\|_F^2.
\label{eq:benchmark_proj_hdm}
\end{equation}
We solve \eqref{eq:benchmark_proj_hdm} using the same ADMM framework to handle constraints \eqref{eq:max_sum_rate_b}--\eqref{eq:max_sum_rate_f}. With fixed $\{(\mathbf P_n^\mathrm{Proj})^\sharp\}$, digital precoders are optimized to obtain $\{\mathbf{W}_n^\mathrm{Proj}\}$ \cite{WMMSE}. Finally, $\{(\mathbf P_n^\mathrm{Proj})^\sharp\}$ and $\{\mathbf{W}_n^\mathrm{Proj}\}$ are plugged into \eqref{eq:max_sum_rate_a}. The average sum-rate is evaluated using the resulting MiLAC and digital beamformers.

\subsubsection{FD-MiLAC} \textbf{FD-MiLAC} scheme uses $N_\mathrm{RF}=K$ MiLAC input ports and jointly optimizes the MiLAC beamformer and digital power allocation. The design problem is
\begin{equation}
\begin{aligned}
&\left(\{(\mathbf P_n^\mathrm{M})^\sharp\},\{(\mathbf{p}^\mathrm{D}_n)^\sharp\}\right)\\
&\in
\operatorname*{\arg~\max}_{\substack{
    \text{\eqref{eq:max_sum_rate_b}-\eqref{eq:max_sum_rate_f}},\\
    \sum_{n=1}^{N}\|\mathbf P_n^\mathrm{M}
       \mathrm{diag}(\mathbf p_n^\mathrm{D})^{\frac{1}{2}}\|_F^2
    \leq P_{\mathrm T}
}}
\frac{1}{N}\sum_{n=1}^{N}R_n\left(
\mathbf P_n^\mathrm{M},\mathrm{diag}(\mathbf p_n^\mathrm{D})^{1/2}
\right).
\end{aligned}
\label{eq:benchmark_fd_milac}
\end{equation}
We solve \eqref{eq:benchmark_fd_milac} using the proposed BCD-ADMM algorithm by setting $\mathbf W_n=\mathrm{diag}(\mathbf p_n^{\mathrm D})^{1/2},~\forall n\in\mathcal{N}$. The average sum-rate is evaluated using the optimized $\{(\mathbf P_n^\mathrm{M})^\sharp\}$ and $\{(\mathbf{p}^\mathrm{D}_n)^\sharp\}$.

\subsubsection{FI-HDA} \textbf{FI-HDA} scheme follows the two-stage conventional phase-shifter-based hybrid digital-analog beamforming design in~\cite{hybrid_ofdm}.\footnote{Frequency-dependent analog beamforming can be realized using additional wideband hardware such as true-time-delay or phase-time networks; such architectures involve different
hardware assumptions and are outside the scope of this comparison. Thus FD-HDA is not studied as a benchmark here.} A common analog precoder is designed using an ideal FI fully connected phase-shifter network.

\subsection{Comparison with Frequency-Independent MiLAC Designs}

\begin{figure}
    \centering
    \includegraphics[width=0.4\textwidth]{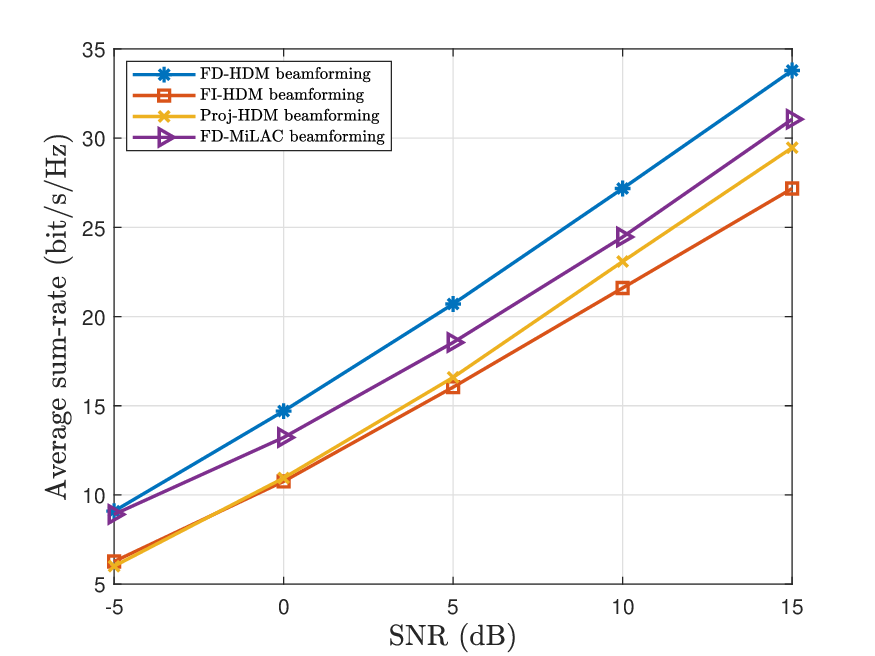}
    \caption{Average sum-rate versus SNR for FD and FI MiLAC-based beamforming schemes ($N_{\mathrm{RF}}=K=4$ and $\tau_{\mathrm{rms}}/T_s=0.75$).}\label{fig:SNR}
\end{figure}

Fig.~\ref{fig:SNR} compares FD-HDM and FD-MiLAC with the FI MiLAC-based benchmark designs versus SNR. FD-HDM consistently achieves the highest sum-rate over the considered SNR range. Its gain over FI-HDM demonstrates the performance loss caused by designing the MiLAC under a frequency-independent assumption, highlighting the importance of directly incorporating the physical frequency-dependent MiLAC circuit model into the optimization. Proj-HDM also exhibits a substantial performance gap, showing that projecting an FI MiLAC design onto the physical FD circuit constraints cannot recover the performance of direct FD optimization. Moreover, FD-MiLAC outperforms both FI-HDM and Proj-HDM despite employing only digital power allocation, further confirming the importance of accounting for MiLAC frequency dependence in the beamforming design.

\subsection{Comparison of Different Beamforming Architectures}

\begin{figure}
    \centering
    \includegraphics[width=0.4\textwidth]{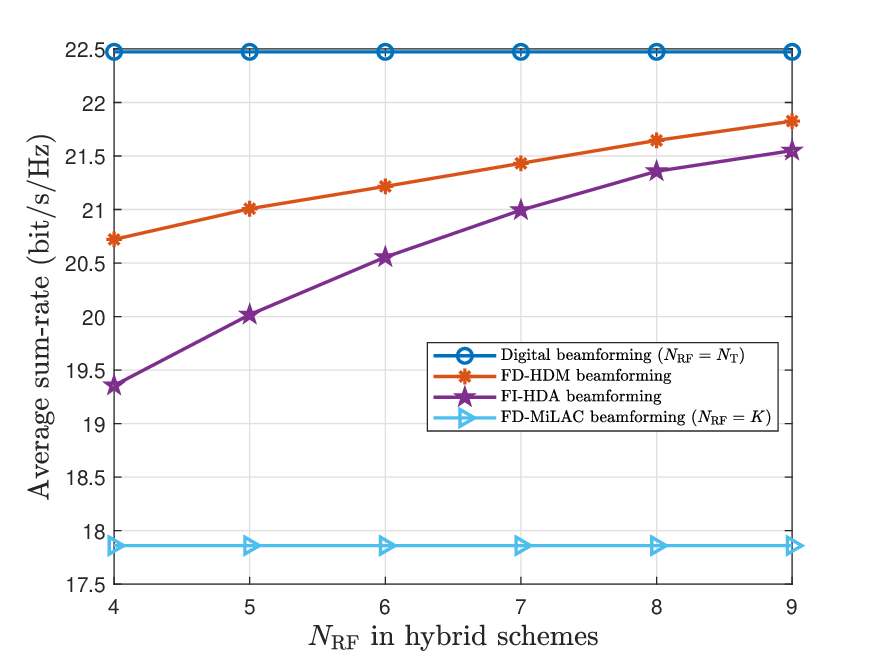}
    \caption{Average sum-rate versus $N_{\mathrm{RF}}$ for different beamforming architectures. Digital and FD-MiLAC beamforming use $N_{\mathrm{RF}}=N_{\mathrm T}=64$ and $N_{\mathrm{RF}}=K=4$, respectively (SNR $=5$ dB, $\tau_{\mathrm{rms}}/T_s=0.75$ and $K=4$).}\label{fig:N_RF}
\end{figure}

\begin{figure}
    \centering
    \includegraphics[width=0.4\textwidth]{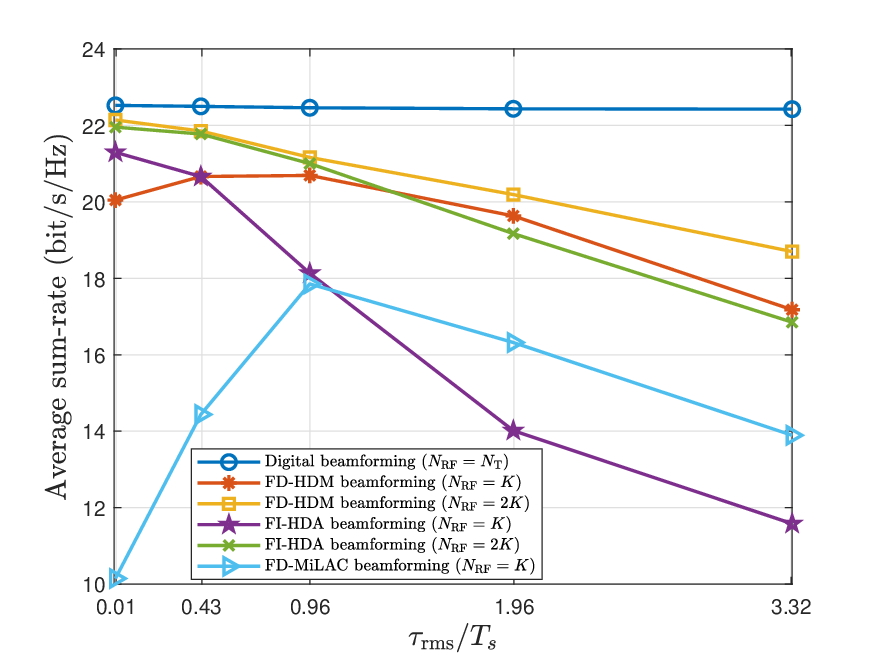}
    \caption{Average sum-rate versus normalized RMS delay spread $\tau_{\mathrm{rms}}/T_s$ for different beamforming architectures (SNR $=5$ dB and $K=4$).}\label{fig:frequency_selectivity}
\end{figure}

Fig.~\ref{fig:N_RF} compares FD-HDM with fully-digital, FI-HDA, and FD-MiLAC beamforming architectures as the number of RF chains varies. With the same $N_{\mathrm{RF}}$ and digital precoding architecture, FD-HDM consistently outperforms FI-HDA, demonstrating the additional cross-subcarrier spatial flexibility provided by the frequency-dependent MiLAC response. The performance gap is more pronounced for a small number of RF chains and gradually narrows as $N_{\mathrm{RF}}$ increases, indicating that this additional flexibility is particularly beneficial when the digital degrees of freedom are limited. At $N_{\mathrm{RF}}=8$, FD-HDM achieves $96.3\%$ of the fully-digital sum-rate while using only $12.5\%$ of the RF chains required by fully-digital beamforming.

Fig.~\ref{fig:frequency_selectivity} illustrates the impact of channel frequency selectivity on the average sum-rate, where $\tau_{\mathrm{rms}}/T_s$ is varied by setting the PDP decay factor to $\epsilon\in\{0.1,0.5,1,2,4\}$. When the channel is nearly frequency-flat, i.e., $\tau_{\mathrm{rms}}/T_s\approx0.01$, FI-HDA remains competitive with FD-HDM; in particular, with $N_{\mathrm{RF}}=K$, FI-HDA outperforms FD-HDM. This is because the desired beamforming direction on each subcarrier is almost identical with a frequency-flat channel. Hence, the FI analog precoder of FI-HDA can provide a suitable spatial subspace over the entire bandwidth. In contrast, due to the inherent frequency dependence of the MiLAC circuit in \eqref{eq:B_linear_model}, a fixed set of circuit parameters generally produces different MiLAC beamformers across subcarriers even when such frequency variation is not required by the channel. As the channel becomes more frequency-selective, the performance gaps of FD-HDM and FI-HDA to digital beamforming increase, while the advantage of FD-HDM over FI-HDA becomes more pronounced. Interestingly, the sum-rate of FD-MiLAC initially increases significantly and peaks around $\tau_{\mathrm{rms}}/T_s\approx0.96$, indicating that the frequency-dependent MiLAC beamformer can beneficially exploit the moderate frequency selectivity of channel even without a full digital precoder. At $\tau_{\mathrm{rms}}/T_s\approx1.96$ and $3.32$, FD-HDM with $N_{\mathrm{RF}}=2K$ achieves $90.1\%$ and $83.5\%$ of the fully-digital sum-rate, respectively, while using only $12.5\%$ of the RF chains required by fully-digital beamforming. Moreover, at $\tau_{\mathrm{rms}}/T_s\approx3.32$, FD-HDM with $N_{\mathrm{RF}}=K$ provides a $48\%$ sum-rate gain over FI-HDA. It demonstrates that the benefit of FD-HDM beamforming becomes increasingly significant as the channel frequency selectivity increases, offering a favorable tradeoff between wideband beamforming performance and the number of RF chains.

\begin{figure}
    \centering
    \includegraphics[width=0.4\textwidth]{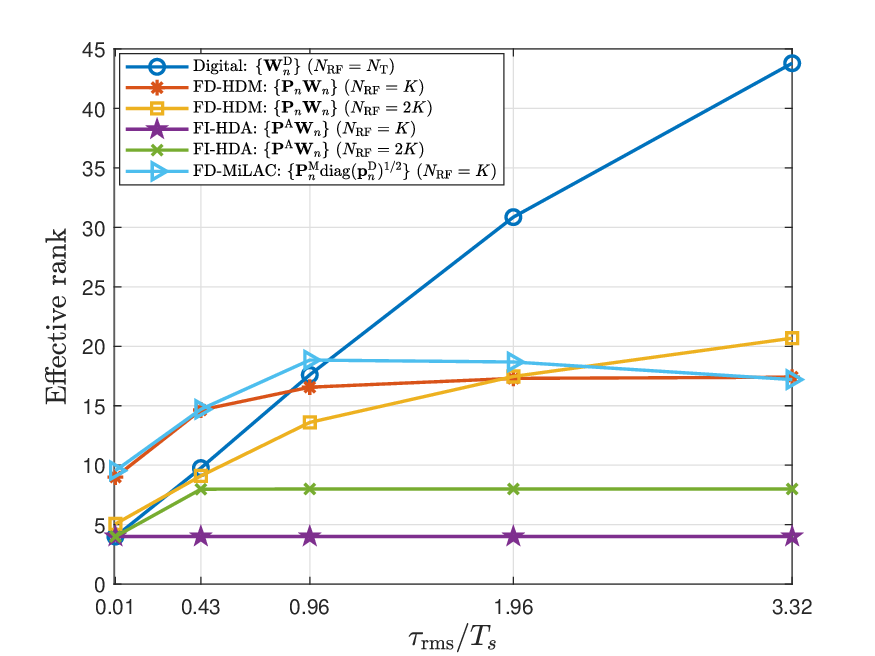}
    \caption{Effective rank of the concatenated transmitted beamformers versus normalized RMS delay spread $\tau_{\mathrm{rms}}/T_s$ for different beamforming architectures (SNR $=5$ dB and $K=4$).}\label{fig:effective_rank}
\end{figure}

To explain these trends, we examine the effective rank of the transmitted beamformers, which characterizes their aggregate spatial dimensionality across subcarriers. For a nonzero matrix $\mathbf X\in\mathbb C^{M_\mathrm{r}\times M_\mathrm{c}}$, let $\sigma_1(\mathbf X)\geq\cdots\geq\sigma_{r_{\max}}(\mathbf X)\geq0$ denote its singular values, where $r_{\max}\triangleq\min\{M_\mathrm{r},M_\mathrm{c}\}$. We define its effective rank as
\begin{equation}
r_{\mathrm{eff}}(\mathbf X)
\triangleq
\min\left\{
1\leq r'\leq r_{\max}~\middle\vert~
\frac{\sum_{i=1}^{r'}\sigma_i^2(\mathbf X)}
{\sum_{i=1}^{r_{\max}}\sigma_i^2(\mathbf X)}
\geq 0.99
\right\}.
\end{equation}
Thus, $r_{\mathrm{eff}}(\mathbf X)$ is the minimum number of singular modes containing at least $99\%$ of the matrix energy. Let the superscripts $\mathrm D$, $\mathrm H$, $\mathrm M$, and $\mathrm A$ denote digital, FD-HDM, FD-MiLAC, and FI-HDA beamforming, respectively, and define $\mathbf X_n^{\mathrm D}\triangleq \frac{1}{2}\mathbf W_n^{\mathrm D},~\mathbf X_n^{\mathrm H}\triangleq \mathbf P_n\mathbf W_n,~\mathbf X_n^{\mathrm M}\triangleq\mathbf P_n^{\mathrm M}\mathrm{diag}(\mathbf p_n^{\mathrm D})^{1/2},~\mathbf X_n^{\mathrm A}\triangleq\mathbf P^{\mathrm A}\mathbf W_n$. Here, $\mathbf P^{\mathrm A}\in\mathbb C^{N_{\mathrm T}\times N_{\mathrm{RF}}}$ denotes the FI phase-shifter analog precoder. For each scheme $\mathrm S\in\{\mathrm D,\mathrm H,\mathrm M,\mathrm A\}$, We concatenate the beamformers across all subcarriers as $\mathbf X^{\mathrm S}\triangleq[\mathbf X_1^{\mathrm S},\mathbf X_2^{\mathrm S},\ldots,\mathbf X_N^{\mathrm S}]\in\C^{N_\mathrm{T}\times KN}$. 

Fig.~\ref{fig:effective_rank} shows the effective rank of the aggregated transmit beamformers for different beamforming architectures, which characterizes the number of dominant spatial dimensions occupied across the bandwidth. At $\tau_{\mathrm{rms}}/T_s\approx0.01$, the fully-digital beamformer has an effective rank of $4$, indicating that nearly frequency-flat channels require only a low-dimensional common spatial subspace. In this regime, FD-HDM with $N_{\mathrm{RF}}=K$ has a higher effective rank of approximately $9$. This higher rank reflects additional cross-subcarrier spatial variation, which spreads the transmit energy over spatial directions that are not required in the nearly frequency-flat regime. This explains why FI-HDA slightly outperforms FD-HDM. With $N_{\mathrm{RF}}=2K$, the effective rank of FD-HDM becomes closer to that of the fully-digital solution due to the additional digital degrees of freedom.

As the channel becomes more frequency-selective, the fully-digital effective rank increases rapidly, reaching $17.6$ at $\tau_{\mathrm{rms}}/T_s\approx0.96$. FI-HDA remains restricted by its frequency-independent analog subspace, whereas FD-HDM can exploit the frequency variation of the MiLAC response to span substantially more spatial dimensions across subcarriers. With $N_{\mathrm{RF}}=K$, the effective rank of FD-HDM reaches $16.56$, which explains the growing performance advantage over FI-HDA. FD-MiLAC reaches an effective rank of $18.84$ at $\tau_{\mathrm{rms}}/T_s\approx0.96$, consistent with its significant sum-rate improvement in the moderately frequency-selective regime. These results show that the frequency dependent MiLAC beamformer provides useful cross-subcarrier spatial flexibility as channel frequency selectivity increases.

\subsection{Effective MiLAC Temporal Memory and CP Overhead}
\begin{figure}
    \centering
    \subfloat[Reconstruction error]{
    \includegraphics[width=0.23\textwidth]{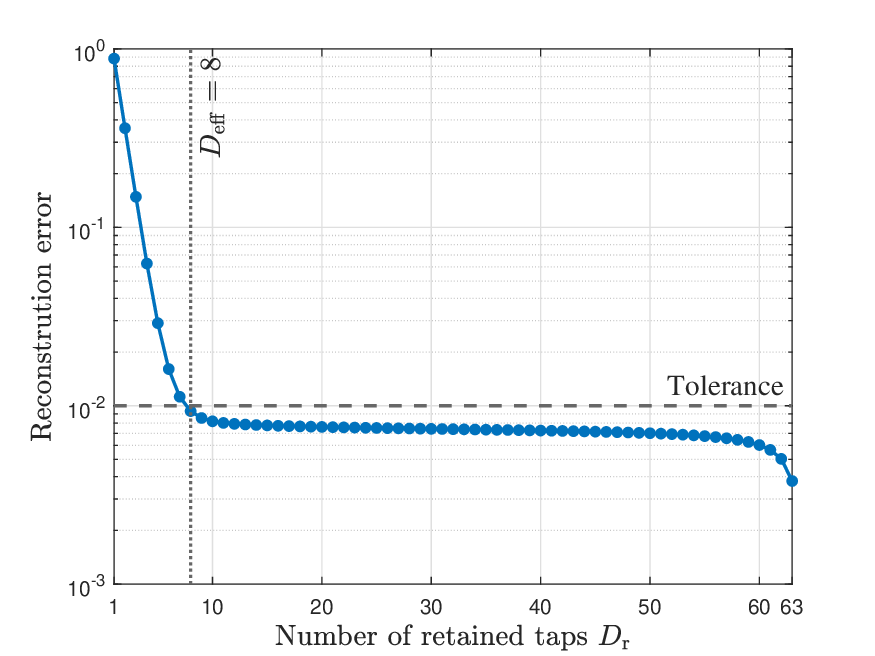}\label{fig:tap_response_error}}
    \subfloat[Average sum-rate]{
    \includegraphics[width=0.23\textwidth]{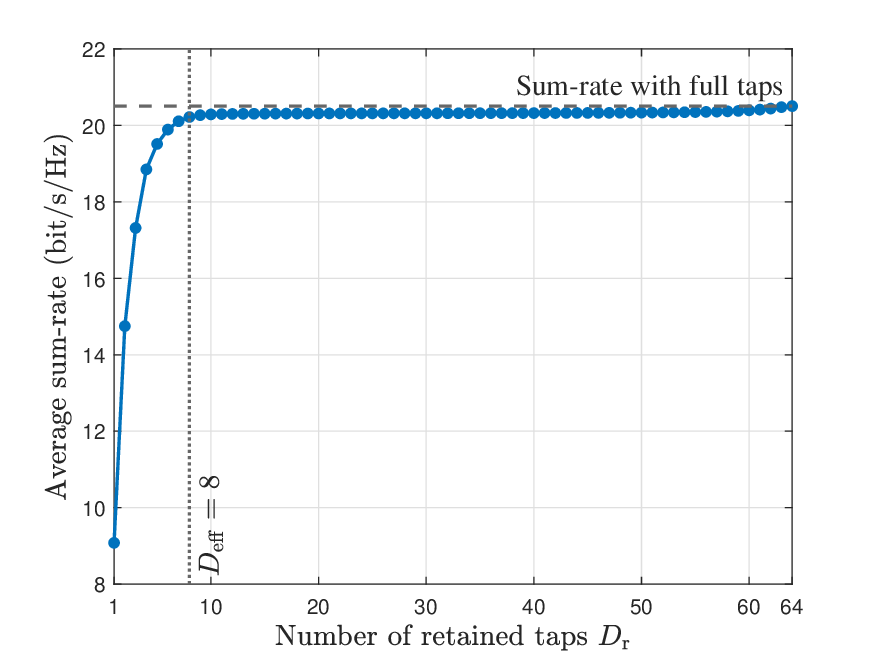}\label{fig:tap_sum_rate}}
    \caption{Characterization of the effective MiLAC temporal memory: (a) reconstruction error and (b) average sum-rate versus the number of retained taps $D_{\mathrm r}$, with $N_{\mathrm{RF}}=K=4$, SNR $=5$ dB, and $\tau_{\mathrm{rms}}/T_s=0.75$.}\label{fig:milac_tap_approximation}
\end{figure}

We next characterize the effective temporal memory introduced by the frequency-dependent MiLAC and evaluate its impact on the required OFDM
CP. For the optimized $\{\mathbf P_n^\star\}$, the corresponding discrete-time
response is obtained by the $N$-point IDFT as $\bar{\mathbf P}_{d'}^\star
=\frac{1}{N}\sum_{n=1}^{N}\mathbf P_n^\star
e^{\jmath\frac{2\pi}{N}(n-1)d'},~d'=0,1,\ldots,N-1$. Retaining the first $D_{\mathrm r}$ taps yields $ \mathbf P_n^{(D_{\mathrm r})}=\sum_{d'=0}^{D_{\mathrm r}-1}\bar{\mathbf P}_{d'}^\star e^{-\jmath\frac{2\pi}{N}(n-1)d'},~n\in\mathcal N$. For each channel realization, we define the reconstruction error as
\begin{equation}
    \mathcal E(D_{\mathrm r})\triangleq
    \frac{\sum_{n=1}^{N}
    \|\mathbf P_n^\star-\mathbf P_n^{(D_{\mathrm r})}\|_F^2}
    {\sum_{n=1}^{N}\|\mathbf P_n^\star\|_F^2}=
    \frac{\sum_{d'=D_{\mathrm r}}^{N-1}
    \|\bar{\mathbf P}_{d'}^\star\|_F^2}
    {\sum_{d'=0}^{N-1}\|\bar{\mathbf P}_{d'}^\star\|_F^2}.
\label{eq:milac_reconstruction_error}
\end{equation}
We define the effective MiLAC memory length as $D_{\mathrm{eff}}
\triangleq
\min\left\{
D_{\mathrm r}\in\{1,\ldots,N\}~\middle\vert~
\mathcal E(D_{\mathrm r})\leq 0.01
\right\}$, i.e., the minimum number of taps containing at least $99\%$ of the response energy. Thus, $D_{\mathrm{eff}}$ characterizes the effective discrete-time temporal spread of the MiLAC response over the OFDM bandwidth. As shown in Fig.~\ref{fig:tap_response_error}, the median effective memory length is $D_{\mathrm{eff}}=8$. Moreover, Fig.~\ref{fig:tap_sum_rate} shows that retaining eight taps preserves $98.6\%$ of the full-response sum-rate. 

We next quantify the corresponding CP requirement. With a $D_{\mathrm{eff}}$-tap MiLAC representation and a $D$-tap wireless channel, the approximated cascaded response has $D+D_{\mathrm{eff}}-1$ taps, and a sufficient CP length is therefore $N_{\mathrm{CP}}\geq D+D_{\mathrm{eff}}-2$. For $D=16$, $D_{\mathrm{eff}}=8$, and $T_s=1/B=10$~ns, choosing $N_{\mathrm{CP}}=22$ gives a CP duration of $T_{\mathrm{CP}}=220$~ns and a useful symbol duration of $T_{\mathrm u}=640$~ns, resulting in a CP overhead of $\frac{T_{\mathrm{CP}}}{T_{\mathrm u}+T_{\mathrm{CP}}}\approx25.6\%$ of the complete OFDM symbol. We further evaluate the CP overhead for $B=200$~MHz and $N\in\{128,256\}$. As shown in Table~\ref{tab:milac_cp_overhead}, $D_{\mathrm{eff}}=12$ in both cases, yielding $N_{\mathrm{CP}}=26$ and $T_{\mathrm{CP}}=130$~ns. The resulting CP overheads decrease to $16.88\%$ and $9.22\%$ for $N=128$ and $256$, respectively. These estimated overheads are below the $20\%$ CP overhead of IEEE~802.11n~\cite{ieee80211n2009}; moreover, the $9.22\%$ overhead for $N=256$ is comparable to the $11.1\%$ overhead of IEEE~802.11ax with a $1.6\,\mu\mathrm{s}$ guard interval \cite{ieee80211ax2021}.

\begin{table}[t]
    \centering
    \caption{Estimated CP overhead at $B=200$~MHz and $\delta=10^{-2}$.}
    \label{tab:milac_cp_overhead}
    \small
    \setlength{\tabcolsep}{5pt}
    \begin{tabular}{ccccc}
        \hline
        $N$ & $D_{\mathrm{eff}}$
        & $T_{\mathrm{CP}}$ (ns)
        & $T_{\mathrm u}$ (ns)
        & CP overhead (\%) \\
        \hline
        128 & 12 & 130 & 640  & 16.88 \\
        256 & 12 & 130 & 1280 & 9.22  \\
        \hline
    \end{tabular}
\end{table}

\section{Conclusions}\label{sec:conclusions}

This paper investigated frequency-dependent hybrid digital-MiLAC beamforming for MU-MISO-OFDM systems. We developed a wideband hybrid digital-MiLAC model that captures the frequency coupling of the MiLAC circuit responses across subcarriers, and showed that a single MiLAC generally cannot simultaneously realize the desired beamforming directions over the entire bandwidth. We then formulated the joint digital and MiLAC beamforming problem and developed a BCD-ADMM algorithm with reduced-complexity computations. 

Simulation results showed that the proposed FD-HDM design approaches fully-digital beamforming performance with substantially fewer RF chains, achieving $96.3\%$ of the fully-digital sum-rate while using only $12.5\%$ of the RF chains. Its advantage over conventional hybrid digital-analog beamforming becomes particularly pronounced under highly frequency-selective channels and small RF-chain numbers; when only $6.25\%$ of the fully-digital RF chains are used, the proposed design provides a $48\%$ sum-rate gain over FI-HDA. Finally, we show that the effective temporal memory of MiLAC can be accommodated with an OFDM CP overhead as low as $9.22\%$.

\bibliographystyle{IEEEtran}
\bibliography{refs}

\end{document}